\documentclass[nofootinbib,pra,twocolumn,showpacs,superscriptaddress,notitlepage,superscriptaddress,amsmath]{revtex4-2} 
\usepackage[colorlinks=true,citecolor=blue,linkcolor=blue,urlcolor=blue]{hyperref}

\usepackage{amsmath,amssymb,bm,color}
\usepackage{dsfont}
\usepackage{framed}
\usepackage{braket}
\definecolor{shadecolor}{gray}{0.95}
\usepackage{amsthm}
\usepackage[normalem]{ulem}

\usepackage{tikz}
\usetikzlibrary{external}
\usetikzlibrary{decorations.pathreplacing}
\usetikzlibrary{decorations.text}

\definecolor{newblue}{RGB}{112,178,255}
\definecolor{neworange}{RGB}{255,204,112}
\definecolor{blue2}{RGB}{120,0,255}
\definecolor{red2}{RGB}{255,0,120}
\definecolor{green2}{RGB}{0,130,130}

\definecolor{darkred}{RGB}{245,186,183}
\definecolor{lightred}{RGB}{249,217,215}

\def\ket#1{\mathinner{|{#1}\rangle}}

\usetikzlibrary{arrows,calc,external,shapes.geometric}

\tikzset{
	>=stealth',
	help lines/.style={dashed, thick},
	important line/.style={thick},
	connection/.style={thick, dotted},
}

\tikzstyle{A}=[circle,draw=red!50,fill=red!20,thick]
\tikzstyle{R}=[circle,draw=blue!50,fill=blue!20,thick]
\tikzstyle{U}=[circle,draw=green!50,fill=green!20,thick]
\tikzstyle{V}=[circle,draw=orange!50,fill=orange!20,thick]

\usepackage{lineno}

\begin{document}

\begin{abstract}
Quantum systems evolve in time in one of two ways: through the Schr\"odinger equation or wavefunction collapse. So far, deterministic control of quantum many-body systems in the lab has focused on the former, due to the probabilistic nature of measurements. This imposes serious limitations: preparing long-range entangled states, for example, requires extensive circuit depth if restricted to unitary dynamics. In this work, we use mid-circuit measurement and feed-forward to implement deterministic non-unitary dynamics on Quantinuum's H1 programmable ion-trap quantum computer. Enabled by these capabilities, we demonstrate for the first time a constant-depth procedure for creating a toric code ground state in real-time. In addition to reaching high stabilizer fidelities, we create a non-Abelian defect whose presence is confirmed by transmuting anyons via braiding. This work clears the way towards creating complex topological orders in the lab and exploring deterministic non-unitary dynamics via measurement and feed-forward.
\end{abstract}

\title{Topological Order from Measurements and Feed-Forward\\on a Trapped Ion Quantum Computer}

\author{Mohsin Iqbal}
\affiliation{Quantinuum, Leopoldstrasse 180, 80804 Munich, Germany}
\author{Nathanan Tantivasadakarn}
\affiliation{Walter Burke Institute for Theoretical Physics and Department of Physics,
California Institute of Technology, Pasadena, CA 91125, USA}

\author{Thomas M. Gatterman}
\affiliation{Quantinuum, 303 S Technology Ct, Broomfield, CO 80021, USA}
\author{Justin A. Gerber}
\affiliation{Quantinuum, 303 S Technology Ct, Broomfield, CO 80021, USA}
\author{Kevin Gilmore}
\affiliation{Quantinuum, 303 S Technology Ct, Broomfield, CO 80021, USA}
\author{Dan Gresh}
\affiliation{Quantinuum, 303 S Technology Ct, Broomfield, CO 80021, USA}
\author{Aaron Hankin}
\affiliation{Quantinuum, 303 S Technology Ct, Broomfield, CO 80021, USA}
\author{Nathan Hewitt}
\affiliation{Quantinuum, 303 S Technology Ct, Broomfield, CO 80021, USA}
\author{Chandler V. Horst}
\affiliation{Quantinuum, 303 S Technology Ct, Broomfield, CO 80021, USA}
\author{Mitchell Matheny}
\affiliation{Quantinuum, 303 S Technology Ct, Broomfield, CO 80021, USA}
\author{Tanner Mengle}
\affiliation{Quantinuum, 303 S Technology Ct, Broomfield, CO 80021, USA}
\author{Brian Neyenhuis}
\affiliation{Quantinuum, 303 S Technology Ct, Broomfield, CO 80021, USA}

\author{Ashvin Vishwanath}
\affiliation{Department of Physics, Harvard University, Cambridge, MA 02138, USA}
\author{Michael Foss-Feig}
\affiliation{Quantinuum, 303 S Technology Ct, Broomfield, CO 80021, USA}
\author{Ruben Verresen}
\affiliation{Department of Physics, Harvard University, Cambridge, MA 02138, USA}
\author{Henrik Dreyer}
\affiliation{Quantinuum, Leopoldstrasse 180, 80804 Munich, Germany}

\date{\today}

\maketitle

Long-range entangled quantum states are central to different branches of modern physics. They appear as error correction codes in quantum information~\cite{dennis2002topological}, emerge as topologically ordered phases in condensed matter, and play a role in lattice gauge theories of high energy physics~\cite{wen_quantum_2010}. Quantum computers and simulators provide new means of exploring such states and tackling their open questions \cite{altman_quantum_2021}.
A number of quantum algorithms have been designed for these devices, many of which can be decomposed into two steps: a \emph{state preparation} step and a \emph{processing} step, in which e.g., unitary dynamics is applied~\cite{kitaev_quantum_1995,lu_algorithms_2021}. For short-range entangled states, the adiabatic theorem guarantees an (approximate) encoding circuit whose depth is independent of the system size. In contrast, long-range entangled states require circuits of \emph{extensive} depth for their preparation~\cite{bravyi_lieb-robinson_2006} due to finite Lieb-Robinson velocities, which bound the spread of correlation in unitary dynamics \cite{lieb_finite_1972}. This situation is problematic: coherence time is a precious resource for near-term quantum computers and simulators and it should not be exhausted during state preparation.

\begin{figure}[!t]
    \centering
\includegraphics[scale=0.8]{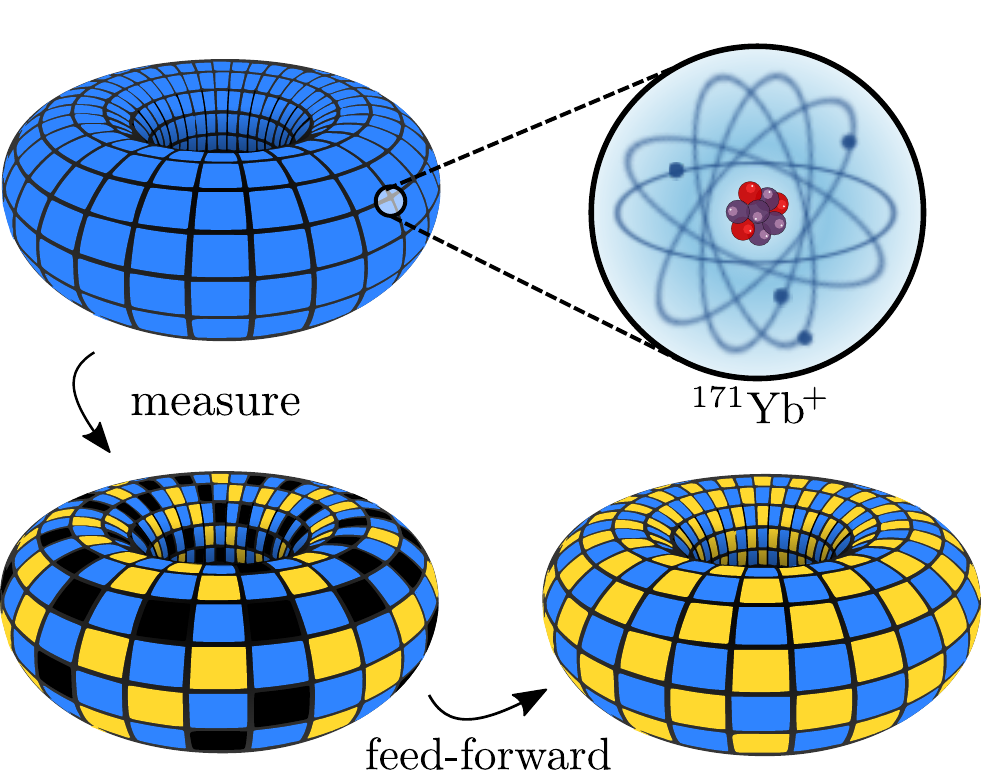}
    \caption{\textbf{Schematic representation of the the toric code ground state from wavefunction collapse.} We initialize a system of trapped-ion qubits (encoded in hyperfine states of $^{171}{\rm Yb}^{+}$) in a product state where all stabilizers $B_p = Z^{\otimes 4}=1$ (blue) are satisfied. We measure $A_p = X^{\otimes 4}$ on every other plaquette, randomly leading to $A_p=1$ (gold) or $A_p=-1$ (black, denoting an $e$-anyon). We use feed-forward to pair up and annihilate the $e$-anyons in real time, deterministically producing a clean toric code wavefunction using a finite-depth circuit and nonlocal classical processing.}
    \label{fig:Fig0}
\end{figure}

\begin{figure*}[!ht]
	\centering
	\includegraphics[width=1.0\textwidth]{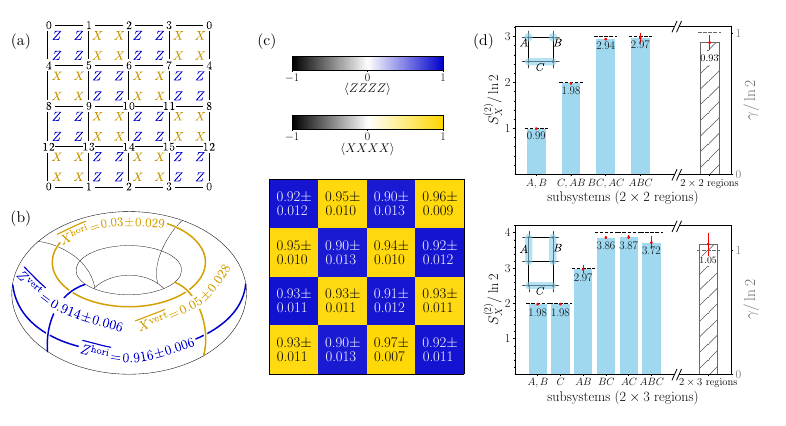}
	\caption{\textbf{Toric code ground state preparation.} (a) Definition of the stabilizer operators~(\ref{eq_toric_code}) on the unraveled torus. Numbers denote the different ions and specify the boundary conditions. Plaquettes are labeled by their upper left qubits. The state comprises 4 $\times$ 4 qubits and periodic boundary conditions.
 (b) Logical $Z$ string operators are $Z^\text{hori} = Z_0 Z_1 Z_2 Z_3$ ($Z^\text{vert} = Z_0 Z_4 Z_8 Z_{12}$) and their vertical (horizontal) translations. $\overline{Z^\text{hori}}$ and $\overline{Z^\text{vert}}$ denote expectation values of the logical string operators, averaged over translations. (c) Expectation values of the stabilizers obtained from the measurement described in the main text. Error bars denote one standard error on the mean. (d) Entanglement entropy measurement on $2 \times 2$ (top) and $2 \times 3$ regions (bottom). Colored bars denote $S^{(2)}_X$ for different subsystems of a region with shapes as shown in the inset. Dashed lines show exact values. The maximum error in the estimates of $S^{(2)}_X$ for $2\times 2$ ($2 \times 3$) regions is $\pm0.056$ ($\pm0.091$). Hatched white bars denote average topological entanglement entropies. \label{fig_toric_code}}
\end{figure*}

Fortunately, there is a loophole to these constraints imposed by unitarity and locality. Introducing measurement during state preparation violates the assumption of unitarity, such that correlations can be generated instantaneously across the whole system. However, since measurements are random, deterministic state preparation requires conditional quantum gates to be applied based on the outcome of the mid-circuit measurement---a capability known as \emph{feed-forward}. In effect, measurement allows one to push all the non-constant depth into the classical channel, which is effectively `free' due to the large speed of light and the comparably much larger cost of quantum gates.
The deterministic preparation of an excitation-free state is important for the quantum simulation of topologically ordered systems with non-error corrected devices, but is also a common prerequisite for quantum error correction protocols that realize universal gate sets \cite{RevModPhys.87.307}. Moreover, feedforward is indispensable for the efficient preparation of certain non-Abelian states involving multiple layers of measurement \cite{verresen_efficiently_2022,tantivasadakarn_long-range_2022,bravyi_adaptive_2022,lu_measurement_2022,hierarchy, piroli_quantum_2021}.

In summary, to prepare long-range entangled states deterministically and in constant (quantum) depth, one requires feed-forward, mid-circuit measurement and entangling gates, all with high fidelity and fast compared to the coherence time of the platform. 
While individual elements of this triad have been demonstrated~\cite{nigg_experimental_2014,cramer_repeated_2016,egan_fault-tolerant_2021,chen_exponential_2021,ryan-anderson_realization_2021,ryan-anderson_implementing_2022,aguado_creation_2008,satzinger_realizing_2021, bluvstein_quantum_2022, andersen_observation_2022, xu_digital_2022, krinner_realizing_2022}, combining all of these ingredients in one platform to deterministically create long-range entangled states has proven elusive since the inception of this idea more than a decade ago~\cite{gottesman_stabilizer_1997,aguado_creation_2008,raussendorf_long-range_2005}.

Here, we demonstrate for the first time the deterministic, high-fidelity preparation of long-range entangled quantum states using a protocol with constant
depth (as conceptually represented in Fig.~\ref{fig:Fig0}), using Quantinuum’s H-series programmable ion-trap quantum computer~\cite{pino_demonstration_2021}. We measure fidelities and entanglement entropies of a toric code ground state on periodic boundaries as well as a model with two non-Abelian Ising defects~\cite{Bombin10Defect,KitaevKong12} which we use to demonstrate anyon transmutation and braiding interferometry.

\section{Toric code preparation with feed-forward}
We target the ground state of Kitaev's toric code Hamiltonian \cite{kitaev_fault-tolerant_2003}. For notational convenience, we represent the qubits as living on the vertices of a square lattice \cite{Wen_plaquette}, described by

\begin{align}
\label{eq_toric_code}
    H = -\sum_{p \in \mathcal{A}} A_p -\sum_{p \in \mathcal{B}} B_p,
\end{align}
where the operators $A = X^{\otimes 4}$ and $B = Z^{\otimes 4}$ act on the four-qubit plaquettes of the square lattice and $\mathcal{A}$ and $\mathcal{B}$ denote the sets of $X$-type and $Z$-type plaquettes (cf. Fig.~\ref{fig_toric_code}(a)). Enabled by the effective all-to-all connectivity of the ion-trap, we implement periodic boundary conditions. 
This Hamiltonian realises $\mathbb{Z}_2$ topological order \cite{Read91,Wen91}, with four ground states---while these all satisfy $\langle A_p \rangle = \langle B_p \rangle = 1$, they can be distinguished by logical string operators that wrap around the torus.
To be specific, we target the unique ground state with logical expectation values $\braket{Z^\text{hori}} = \braket{Z^\text{vert}} = 1$, as defined in Fig.~\ref{fig_toric_code}. The expectation value of the string and commuting plaquette operators certify the quality of the state preparation, with an average of +1 indicating perfect ground state preparation.

To prepare the ground state deterministically and in constant depth, we use a three-step procedure: First, all ions are initialised in $\ket{0}$, such that $\braket{B_p} = 1$. Second, we measure the $A_p$ operator on all odd plaquettes, effectively implementing the projectors $(\mathbb{I} \pm X^{\otimes 4})/2$ with equal probability. This can be done with or without ancillae and we choose to demonstrate both strategies, preparing plaquettes 4, 6, 12 and 14 with an ancilla-free procedure while the measurements on plaquettes 1, 3, 9 and 11 are performed with one ancilla each (Methods, see also Fig.~\ref{fig_toric_code} for our labeling convention). Finally, we apply conditional single-qubit $Z$ gates to flip all plaquettes at which $A_p = -1$ has been measured. To find the location at which the conditional $Z$-gates must be applied, we use a simple lookup-table decoder (Methods).

The topology of the toric code requires anyonic defects to come in pairs; however, errors in the syndrome measurement process can result in measuring an odd number of excitations. Therefore, we employ a state preparation strategy---common to many quantum error correction or repeat-until-success protocols---in which odd defect numbers are heralded and the associated data is discarded. We note that unlike post-selection on each plaquette individually, these errors are heralded and even for a completely depolarized state only half of the data would be discarded; thus we can view the discarding of erroneous runs as a scalable part of the state preparation procedure itself, and we report fidelities constructed from the retained data in the main text (see Extended Data Figure~\ref{ext_fig_comparisons} for the raw data).

We test the quality of the state prepared in the above manner in two ways. First, we measure the expectation values of the $X^{\otimes 4}$ and $Z^{\otimes 4}$ stabilizers. Their average plays the role of the energy density of~(\ref{eq_toric_code}) and is closely related to the overlap with the ground state manifold \cite{cramer_2010}. We report an energy density of $-0.929 \pm 0.004$, indicating that a ground state has been prepared with high fidelity. The expectation value of the two logical string operators averaged over translations is close to 1 and equal up to statistical fluctuations, $\overline{Z^\text{hori}} =  0.916 \pm 0.0065$, $\overline{Z^\text{vert}} = 0.914 \pm 0.0064$, indicating that the target logical state is indeed responsible for the bulk of the overlap with the ground space manifold. The average expectation of the $X$-type plaquettes $\braket{A_p} = 0.944 \pm 0.0049$ exceeds that of the $Z$-type plaquettes $\braket{B_p}= 0.914 \pm 0.0063$. This is compatible with the fact that the two-qubit gate noise in the device is known to be slightly biased towards $Z$-type phase flips: The only two-qubit gates in the circuits occur during the measurement of the $X^{\otimes 4}$-operator. Any $Z$-errors that occur on the data qubits during the ancilla-based measurement circuit are transformed into $X$-error by the Hadamard gates at the end of the subroutine. In turn, these bit flip errors corrupt the neighbouring $Z$-type plaquettes, while being invisible to the $X^{\otimes 4}$-operators (cf. Extended Data Figures~\ref{ext_fig_noise_bias} and~\ref{ext_fig_measurement}(a)). To arrive at these numbers, we executed 1240 repetitions of the state preparation procedure of which roughly $10\%$ were discarded via the heralded state-preparation procedure.

\begin{figure*}[!ht]
	\centering
	\includegraphics[width=1.0\textwidth]{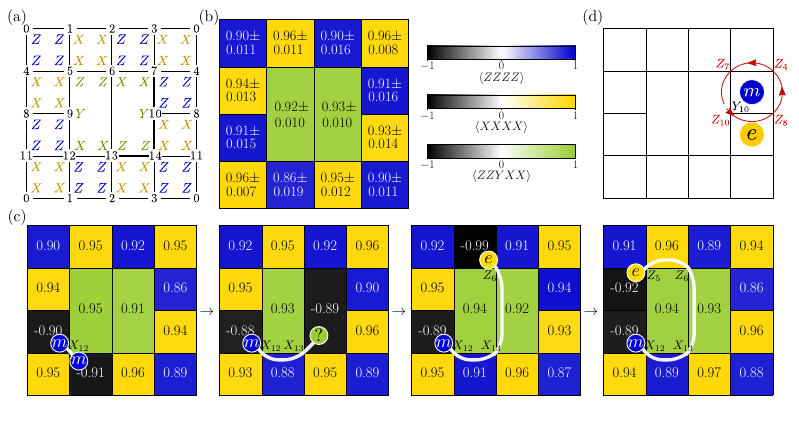}
	\caption{\textbf{Anyon dynamics on a state with two non-Abelian defects.} (a) Geometry. The lack of the central qubit and the redefinition of the stabilizers leads to two defective plaquettes. (b) Expectation values of the stabilizers obtained from the state preparation and measurement routine described in the main text. Error bars denote one standard error on the mean. (c) Anyon transmutation. A pair of magnetic anyons is created and one partner is transmuted into an electric anyon by moving it across the line connecting the two defects. The maximum and minimum error in the expectation values of stabilizers are $\pm0.023$ and $\pm0.0066$ respectively. (d) Anyon Interferometry. A fermionic $e-m$ composite anyon is created next to the defect and a controlled-$Z_{10} Z_{8} Z_{4} Z_{7}$ braiding operations is applied with the help of an ancilla.}
 \label{fig_defect}
\end{figure*}

A second test for the quality of the state is the topological entanglement entropy \cite{kitaev_topological_2006,levin_detecting_2006}. For short-range entangled phases, order parameters can usually be defined in terms of local linear functionals of the density matrix. For long-range entangled states, by definition, no such observables exist. Instead, it is customary to partition a region into areas $A$, $B$ and $C$ and compute the topological entanglement entropy by measuring $\gamma = -(S_A + S_B + S_C - S_{AB} - S_{AC} - S_{BC} + S_{ABC})$ where $S_X$ is the von-Neumann entropy of the reduced density matrix of subsystem $X$. In a phase with $\mathbb{Z}_2$-topological order, $\gamma = \ln 2$ for all Rényi entropies \cite{flammia_topological_2009}. Due to their non-linear nature, entanglement entropies are expensive to measure in practice, requiring a number of shots that is exponential in the size of the subsystem. Here, we employ the randomized measurement scheme~\cite{van2011power, elben_renyi_2018, vermersch_unitary_2018,brydges_probing_2019} to measure $\gamma$ for connected regions of up to six qubits by computing the second Rényi entropies of their subsystems as shown in Fig.~\ref{fig_toric_code}(d) (Methods). We report average topological entanglement entropies of $\gamma/\ln 2 = 0.93 \pm 0.055$  and $\gamma /\ln 2 = 1.05 \pm 0.093$ for the $2 \times 2$ and $2 \times 3$ regions, respectively, indicating that a 
 state consistent with $\mathbb{Z}_2$ topological order has been prepared.

\section{Anyon transmutation and interferometry}

Having established a deterministic procedure to prepare toric code ground states at constant depth with high fidelity, we are now in a position to study simple dynamics on top of the ground state. To this end, we consider a slightly modified geometry, introducing two defects into the system (Fig.~\ref{fig_defect}(a))~\cite{Bombin10Defect,KitaevKong12}. This defective state is only slightly harder to prepare than the toric code ground state (Methods), and we report an average expectation value per plaquette of $0.925 \pm 0.0039$ (Fig.~\ref{fig_defect}(b)), comparable to the toric code ground state considered before. The geometry with defects lends itself to the study of two types of dynamics.

In the first experiment, we study the transmutation of anyons. The elementary excitations of the defect-free toric code are electric and magnetic anyons (corresponding to violations of stabilizers on $X$-type and $Z$-type plaquettes, respectively) as well as their bound state. While such particles can be moved diagonally through the system and annihilated in pairs, their type is fixed throughout the evolution. The insertion of a defect changes this situation: Moving an anyon across the line connecting the two defective plaquettes allows the particle to \emph{skip} a square, moving between $X$-type and $Z$-type plaquettes, and thus change its nature. We choose to create a pair of magnetic particles and move one of them across the defect on the path shown in (Fig.~\ref{fig_defect}(c)), performing a measurement on all qubits after each step. We report final stabilizer expectation values of $-0.92\pm 0.017$ and $-0.89\pm 0.020$ on adjacent plaquettes. The creation of a \emph{single} electric-magnetic pair is impossible in a defect-free toric code and is related to the non-Abelian nature of the defect.
Indeed, such an $e-m$ composite is a fermion (due to the mutual statistics of $e$ and $m$ anyons), and the defect can be thought of as a Majorana zero mode whose fermion parity can be toggled by pulling out a single fermion~\cite{you_synthetic_2013}.
The data indicates that the creation and movement of the anyons does not affect the bystanding plaquettes beyond statistical fluctuation, showing that cross-talk is negligible as it is expected from a quantum charge coupled device in which ions are stored $\geq 180 \mu$m apart. In principle, we can also trace the anyon non-destructively by performing parity measurements instead of collapsing the full wavefunction at every step. This procedure reduces the required number of shots by a factor that is equal to the number of steps, at the cost of introducing extra gates. The results of this strategy are reported in Extended Data Figure~\ref{ext_fig_nd_trans}.

As shown in the transmutation experiment, the presence of the non-Abelian defect allows for the creation of a \emph{single} fermionic excitation (in the form of an $e-m$ composite). 
Here we explicitly confirm that we have created a fermion by checking that its wavefunction picks up a minus sign upon rotating it by 360$^\circ$.
Equivalently, the two anyons making up the composite have non-trivial mutual braiding:
the wavefunction acquires a global phase $U_\text{braid}\ket{em} = -\ket{em}$ when braiding one particle around the other. This phase is naively inaccessible, but it can be measured using the Hadamard test: A controlled version of $U_\text{braid}$ is applied, conditioned on the state of an ancilla which is initially prepared in $\ket{+} = (\ket{0} + \ket{1})/\sqrt{2}$. The phase $\braket{em|U_\text{braid}|em}$ is then directly related to the expectation value $\braket{X}$ on the ancilla. Specifically, in the experiment, we create a single electric-magnetic pair adjacent to the defect by acting with $Y_{10}$ on the ground state (i.e., $\ket{em}:=Y_{10}\ket{\text{gs}}$) and braid the electric around the magnetic excitation on the path shown in Fig.~\ref{fig_defect}(d). We find $\Re\braket{em|U_\text{braid}|em} = -0.87 \pm 0.018$, verifying the fermionic exchange statistics. Similarly, in the absence of the fermionic excitation, we find $\Re\braket{\text{gs}|U_\text{braid}|\text{gs}} = +0.87 \pm 0.018$. The strength of the interferometric signal is remarkable: While conditional dynamics usually requires the use of many SWAP gates to bring the ancilla close to the target qubits, we achieve the same effect here with only four two-qubit gates, due to the effective all-to-all connectivity of the device.

\section{Discussion and Outlook}
We have demonstrated the combined use of mid-circuit measurement, feed-forward and low-error gates to prepare topologically ordered states deterministically, in constant depth and with high fidelity, providing experimental data from Quantinuum's H1-1 programmable ion-trap quantum computer~\cite{h11}. Furthermore, the effective all-to-all connectivity of the device was vital for the implementation of the periodic two-dimensional geometry and the conditional dynamics.

In this work, we have considered the efficient preparation of Abelian topological orders and of defects with non-Abelian character within the Abelian phases. This lays the groundwork for the preparation of true non-Abelian topological orders in a quantum device, some of which, surprisingly, require the same overhead as their Abelian counterparts despite their richer properties~\cite{tantivasadakarn_shortest_2022}. Namely, only a single round of feed-forward and low-error gates suffices to deterministically prepare such states. Moreover, multiple rounds of measurements and feed-forward can access even more exotic and powerful non-Abelian states \cite{verresen_efficiently_2022,tantivasadakarn_long-range_2022,bravyi_adaptive_2022,lu_measurement_2022,hierarchy} which open up new avenues for fault tolerant quantum information processing.

Another line of research concerns the study of more complex dynamics. While the present work considered discrete transformations between eigenstates of the system, our demonstration opens up the possibility of studying quenches and variational circuits in topologically ordered systems on digital quantum computers with minimal resources \cite{heyl_dynamical_2018,cerezo2021variational}. These, in turn, can be used to study, e.g., lattice gauge theories at finite temperatures and energies~\cite{lu_algorithms_2021}. Since classical simulation of these problems generically requires exponential resources, there is a potential for quantum advantage. While the device noise is small and its characterisation agrees well with our experimental findings, the capacity of the H1-1 system must be extended beyond 20 qubits for this to become a reality.

In conclusion, this work has demonstrated a powerful application of measurements and feed-forward. These capabilities open up a multitude of directions for further exploration, ranging from quantum information processing and simulating the ground states and dynamics of many-body quantum systems, to uncovering the emergent structures in monitored circuits \cite{Potter_2022,Fisher22}.






\bibliography{references_henrik_zotero_wen16, other_biblio}

\begin{thebibliography}{58}%
\makeatletter
\providecommand \@ifxundefined [1]{%
 \@ifx{#1\undefined}
}%
\providecommand \@ifnum [1]{%
 \ifnum #1\expandafter \@firstoftwo
 \else \expandafter \@secondoftwo
 \fi
}%
\providecommand \@ifx [1]{%
 \ifx #1\expandafter \@firstoftwo
 \else \expandafter \@secondoftwo
 \fi
}%
\providecommand \natexlab [1]{#1}%
\providecommand \enquote  [1]{``#1''}%
\providecommand \bibnamefont  [1]{#1}%
\providecommand \bibfnamefont [1]{#1}%
\providecommand \citenamefont [1]{#1}%
\providecommand \href@noop [0]{\@secondoftwo}%
\providecommand \href [0]{\begingroup \@sanitize@url \@href}%
\providecommand \@href[1]{\@@startlink{#1}\@@href}%
\providecommand \@@href[1]{\endgroup#1\@@endlink}%
\providecommand \@sanitize@url [0]{\catcode `\\12\catcode `\$12\catcode
  `\&12\catcode `\#12\catcode `\^12\catcode `\_12\catcode `\%12\relax}%
\providecommand \@@startlink[1]{}%
\providecommand \@@endlink[0]{}%
\providecommand \url  [0]{\begingroup\@sanitize@url \@url }%
\providecommand \@url [1]{\endgroup\@href {#1}{\urlprefix }}%
\providecommand \urlprefix  [0]{URL }%
\providecommand \Eprint [0]{\href }%
\providecommand \doibase [0]{https://doi.org/}%
\providecommand \selectlanguage [0]{\@gobble}%
\providecommand \bibinfo  [0]{\@secondoftwo}%
\providecommand \bibfield  [0]{\@secondoftwo}%
\providecommand \translation [1]{[#1]}%
\providecommand \BibitemOpen [0]{}%
\providecommand \bibitemStop [0]{}%
\providecommand \bibitemNoStop [0]{.\EOS\space}%
\providecommand \EOS [0]{\spacefactor3000\relax}%
\providecommand \BibitemShut  [1]{\csname bibitem#1\endcsname}%
\let\auto@bib@innerbib\@empty
\bibitem [{\citenamefont {Dennis}\ \emph {et~al.}(2002)\citenamefont {Dennis},
  \citenamefont {Kitaev}, \citenamefont {Landahl},\ and\ \citenamefont
  {Preskill}}]{dennis2002topological}%
  \BibitemOpen
  \bibfield  {author} {\bibinfo {author} {\bibfnamefont {E.}~\bibnamefont
  {Dennis}}, \bibinfo {author} {\bibfnamefont {A.}~\bibnamefont {Kitaev}},
  \bibinfo {author} {\bibfnamefont {A.}~\bibnamefont {Landahl}},\ and\ \bibinfo
  {author} {\bibfnamefont {J.}~\bibnamefont {Preskill}},\ }\bibfield  {title}
  {\bibinfo {title} {Topological quantum memory},\ }\href@noop {} {\bibfield
  {journal} {\bibinfo  {journal} {Journal of Mathematical Physics}\ }\textbf
  {\bibinfo {volume} {43}},\ \bibinfo {pages} {4452} (\bibinfo {year}
  {2002})}\BibitemShut {NoStop}%
\bibitem [{\citenamefont {Wen}(2010)}]{wen_quantum_2010}%
  \BibitemOpen
  \bibfield  {author} {\bibinfo {author} {\bibfnamefont {X.-G.}\ \bibnamefont
  {Wen}},\ }\href@noop {} {\emph {\bibinfo {title} {Quantum field theory of
  many-body systems}}},\ Oxford {Graduate} {Texts}\ (\bibinfo  {publisher}
  {Oxford University Press},\ \bibinfo {address} {Oxford},\ \bibinfo {year}
  {2010})\BibitemShut {NoStop}%
\bibitem [{\citenamefont {Altman}\ \emph {et~al.}(2021)\citenamefont {Altman},
  \citenamefont {Brown}, \citenamefont {Carleo}, \citenamefont {Carr},
  \citenamefont {Demler}, \citenamefont {Chin}, \citenamefont {DeMarco},
  \citenamefont {Economou}, \citenamefont {Eriksson}, \citenamefont {Fu},
  \citenamefont {Greiner}, \citenamefont {Hazzard}, \citenamefont {Hulet},
  \citenamefont {Kollar}, \citenamefont {Lev}, \citenamefont {Lukin},
  \citenamefont {Ma}, \citenamefont {Mi}, \citenamefont {Misra}, \citenamefont
  {Monroe}, \citenamefont {Murch}, \citenamefont {Nazario}, \citenamefont {Ni},
  \citenamefont {Potter}, \citenamefont {Roushan}, \citenamefont {Saffman},
  \citenamefont {Schleier-Smith}, \citenamefont {Siddiqi}, \citenamefont
  {Simmonds}, \citenamefont {Singh}, \citenamefont {Spielman}, \citenamefont
  {Temme}, \citenamefont {Weiss}, \citenamefont {Vuckovic}, \citenamefont
  {Vuletic}, \citenamefont {Ye},\ and\ \citenamefont
  {Zwierlein}}]{altman_quantum_2021}%
  \BibitemOpen
  \bibfield  {author} {\bibinfo {author} {\bibfnamefont {E.}~\bibnamefont
  {Altman}}, \bibinfo {author} {\bibfnamefont {K.~R.}\ \bibnamefont {Brown}},
  \bibinfo {author} {\bibfnamefont {G.}~\bibnamefont {Carleo}}, \bibinfo
  {author} {\bibfnamefont {L.~D.}\ \bibnamefont {Carr}}, \bibinfo {author}
  {\bibfnamefont {E.}~\bibnamefont {Demler}}, \bibinfo {author} {\bibfnamefont
  {C.}~\bibnamefont {Chin}}, \bibinfo {author} {\bibfnamefont {B.}~\bibnamefont
  {DeMarco}}, \bibinfo {author} {\bibfnamefont {S.~E.}\ \bibnamefont
  {Economou}}, \bibinfo {author} {\bibfnamefont {M.~A.}\ \bibnamefont
  {Eriksson}}, \bibinfo {author} {\bibfnamefont {K.-M.~C.}\ \bibnamefont {Fu}},
  \bibinfo {author} {\bibfnamefont {M.}~\bibnamefont {Greiner}}, \bibinfo
  {author} {\bibfnamefont {K.~R.~A.}\ \bibnamefont {Hazzard}}, \bibinfo
  {author} {\bibfnamefont {R.~G.}\ \bibnamefont {Hulet}}, \bibinfo {author}
  {\bibfnamefont {A.~J.}\ \bibnamefont {Kollar}}, \bibinfo {author}
  {\bibfnamefont {B.~L.}\ \bibnamefont {Lev}}, \bibinfo {author} {\bibfnamefont
  {M.~D.}\ \bibnamefont {Lukin}}, \bibinfo {author} {\bibfnamefont
  {R.}~\bibnamefont {Ma}}, \bibinfo {author} {\bibfnamefont {X.}~\bibnamefont
  {Mi}}, \bibinfo {author} {\bibfnamefont {S.}~\bibnamefont {Misra}}, \bibinfo
  {author} {\bibfnamefont {C.}~\bibnamefont {Monroe}}, \bibinfo {author}
  {\bibfnamefont {K.}~\bibnamefont {Murch}}, \bibinfo {author} {\bibfnamefont
  {Z.}~\bibnamefont {Nazario}}, \bibinfo {author} {\bibfnamefont {K.-K.}\
  \bibnamefont {Ni}}, \bibinfo {author} {\bibfnamefont {A.~C.}\ \bibnamefont
  {Potter}}, \bibinfo {author} {\bibfnamefont {P.}~\bibnamefont {Roushan}},
  \bibinfo {author} {\bibfnamefont {M.}~\bibnamefont {Saffman}}, \bibinfo
  {author} {\bibfnamefont {M.}~\bibnamefont {Schleier-Smith}}, \bibinfo
  {author} {\bibfnamefont {I.}~\bibnamefont {Siddiqi}}, \bibinfo {author}
  {\bibfnamefont {R.}~\bibnamefont {Simmonds}}, \bibinfo {author}
  {\bibfnamefont {M.}~\bibnamefont {Singh}}, \bibinfo {author} {\bibfnamefont
  {I.~B.}\ \bibnamefont {Spielman}}, \bibinfo {author} {\bibfnamefont
  {K.}~\bibnamefont {Temme}}, \bibinfo {author} {\bibfnamefont {D.~S.}\
  \bibnamefont {Weiss}}, \bibinfo {author} {\bibfnamefont {J.}~\bibnamefont
  {Vuckovic}}, \bibinfo {author} {\bibfnamefont {V.}~\bibnamefont {Vuletic}},
  \bibinfo {author} {\bibfnamefont {J.}~\bibnamefont {Ye}},\ and\ \bibinfo
  {author} {\bibfnamefont {M.}~\bibnamefont {Zwierlein}},\ }\bibfield  {title}
  {\bibinfo {title} {Quantum {Simulators}: {Architectures} and
  {Opportunities}},\ }\href {https://doi.org/10.1103/PRXQuantum.2.017003}
  {\bibfield  {journal} {\bibinfo  {journal} {PRX Quantum}\ }\textbf {\bibinfo
  {volume} {2}},\ \bibinfo {pages} {017003} (\bibinfo {year}
  {2021})}\BibitemShut {NoStop}%
\bibitem [{\citenamefont {Kitaev}(1995)}]{kitaev_quantum_1995}%
  \BibitemOpen
  \bibfield  {author} {\bibinfo {author} {\bibfnamefont {A.~Y.}\ \bibnamefont
  {Kitaev}},\ }\href {https://doi.org/10.48550/arXiv.quant-ph/9511026}
  {\bibinfo {title} {Quantum measurements and the {Abelian} {Stabilizer}
  {Problem}}} (\bibinfo {year} {1995})\BibitemShut {NoStop}%
\bibitem [{\citenamefont {Lu}\ \emph {et~al.}(2021)\citenamefont {Lu},
  \citenamefont {Bañuls},\ and\ \citenamefont {Cirac}}]{lu_algorithms_2021}%
  \BibitemOpen
  \bibfield  {author} {\bibinfo {author} {\bibfnamefont {S.}~\bibnamefont
  {Lu}}, \bibinfo {author} {\bibfnamefont {M.~C.}\ \bibnamefont {Bañuls}},\
  and\ \bibinfo {author} {\bibfnamefont {J.~I.}\ \bibnamefont {Cirac}},\
  }\bibfield  {title} {\bibinfo {title} {Algorithms for quantum simulation at
  finite energies},\ }\href {https://doi.org/10.1103/PRXQuantum.2.020321}
  {\bibfield  {journal} {\bibinfo  {journal} {PRX Quantum}\ }\textbf {\bibinfo
  {volume} {2}},\ \bibinfo {pages} {020321} (\bibinfo {year}
  {2021})}\BibitemShut {NoStop}%
\bibitem [{\citenamefont {Bravyi}\ \emph {et~al.}(2006)\citenamefont {Bravyi},
  \citenamefont {Hastings},\ and\ \citenamefont
  {Verstraete}}]{bravyi_lieb-robinson_2006}%
  \BibitemOpen
  \bibfield  {author} {\bibinfo {author} {\bibfnamefont {S.}~\bibnamefont
  {Bravyi}}, \bibinfo {author} {\bibfnamefont {M.~B.}\ \bibnamefont
  {Hastings}},\ and\ \bibinfo {author} {\bibfnamefont {F.}~\bibnamefont
  {Verstraete}},\ }\bibfield  {title} {\bibinfo {title} {Lieb-{Robinson}
  {Bounds} and the {Generation} of {Correlations} and {Topological} {Quantum}
  {Order}},\ }\href {https://doi.org/10.1103/PhysRevLett.97.050401} {\bibfield
  {journal} {\bibinfo  {journal} {Physical Review Letters}\ }\textbf {\bibinfo
  {volume} {97}},\ \bibinfo {pages} {050401} (\bibinfo {year}
  {2006})}\BibitemShut {NoStop}%
\bibitem [{\citenamefont {Lieb}\ and\ \citenamefont
  {Robinson}(1972)}]{lieb_finite_1972}%
  \BibitemOpen
  \bibfield  {author} {\bibinfo {author} {\bibfnamefont {E.~H.}\ \bibnamefont
  {Lieb}}\ and\ \bibinfo {author} {\bibfnamefont {D.~W.}\ \bibnamefont
  {Robinson}},\ }\bibfield  {title} {\bibinfo {title} {The finite group
  velocity of quantum spin systems},\ }\href
  {https://doi.org/10.1007/BF01645779} {\bibfield  {journal} {\bibinfo
  {journal} {Communications in Mathematical Physics}\ }\textbf {\bibinfo
  {volume} {28}},\ \bibinfo {pages} {251} (\bibinfo {year} {1972})}\BibitemShut
  {NoStop}%
\bibitem [{\citenamefont {Terhal}(2015)}]{RevModPhys.87.307}%
  \BibitemOpen
  \bibfield  {author} {\bibinfo {author} {\bibfnamefont {B.~M.}\ \bibnamefont
  {Terhal}},\ }\bibfield  {title} {\bibinfo {title} {Quantum error correction
  for quantum memories},\ }\href {https://doi.org/10.1103/RevModPhys.87.307}
  {\bibfield  {journal} {\bibinfo  {journal} {Rev. Mod. Phys.}\ }\textbf
  {\bibinfo {volume} {87}},\ \bibinfo {pages} {307} (\bibinfo {year}
  {2015})}\BibitemShut {NoStop}%
\bibitem [{\citenamefont {Verresen}\ \emph {et~al.}(2022)\citenamefont
  {Verresen}, \citenamefont {Tantivasadakarn},\ and\ \citenamefont
  {Vishwanath}}]{verresen_efficiently_2022}%
  \BibitemOpen
  \bibfield  {author} {\bibinfo {author} {\bibfnamefont {R.}~\bibnamefont
  {Verresen}}, \bibinfo {author} {\bibfnamefont {N.}~\bibnamefont
  {Tantivasadakarn}},\ and\ \bibinfo {author} {\bibfnamefont {A.}~\bibnamefont
  {Vishwanath}},\ }\href {https://doi.org/10.48550/arXiv.2112.03061} {\bibinfo
  {title} {Efficiently preparing {Schrödinger}'s cat, fractons and
  non-{Abelian} topological order in quantum devices}} (\bibinfo {year}
  {2022})\BibitemShut {NoStop}%
\bibitem [{\citenamefont {Tantivasadakarn}\ \emph
  {et~al.}(2022{\natexlab{a}})\citenamefont {Tantivasadakarn}, \citenamefont
  {Thorngren}, \citenamefont {Vishwanath},\ and\ \citenamefont
  {Verresen}}]{tantivasadakarn_long-range_2022}%
  \BibitemOpen
  \bibfield  {author} {\bibinfo {author} {\bibfnamefont {N.}~\bibnamefont
  {Tantivasadakarn}}, \bibinfo {author} {\bibfnamefont {R.}~\bibnamefont
  {Thorngren}}, \bibinfo {author} {\bibfnamefont {A.}~\bibnamefont
  {Vishwanath}},\ and\ \bibinfo {author} {\bibfnamefont {R.}~\bibnamefont
  {Verresen}},\ }\href {https://doi.org/10.48550/arXiv.2112.01519} {\bibinfo
  {title} {Long-range entanglement from measuring symmetry-protected
  topological phases}} (\bibinfo {year} {2022}{\natexlab{a}})\BibitemShut
  {NoStop}%
\bibitem [{\citenamefont {Bravyi}\ \emph {et~al.}(2022)\citenamefont {Bravyi},
  \citenamefont {Kim}, \citenamefont {Kliesch},\ and\ \citenamefont
  {Koenig}}]{bravyi_adaptive_2022}%
  \BibitemOpen
  \bibfield  {author} {\bibinfo {author} {\bibfnamefont {S.}~\bibnamefont
  {Bravyi}}, \bibinfo {author} {\bibfnamefont {I.}~\bibnamefont {Kim}},
  \bibinfo {author} {\bibfnamefont {A.}~\bibnamefont {Kliesch}},\ and\ \bibinfo
  {author} {\bibfnamefont {R.}~\bibnamefont {Koenig}},\ }\href
  {https://doi.org/10.48550/arXiv.2205.01933} {\bibinfo {title} {Adaptive
  constant-depth circuits for manipulating non-abelian anyons}} (\bibinfo
  {year} {2022})\BibitemShut {NoStop}%
\bibitem [{\citenamefont {Lu}\ \emph {et~al.}(2022)\citenamefont {Lu},
  \citenamefont {Lessa}, \citenamefont {Kim},\ and\ \citenamefont
  {Hsieh}}]{lu_measurement_2022}%
  \BibitemOpen
  \bibfield  {author} {\bibinfo {author} {\bibfnamefont {T.-C.}\ \bibnamefont
  {Lu}}, \bibinfo {author} {\bibfnamefont {L.~A.}\ \bibnamefont {Lessa}},
  \bibinfo {author} {\bibfnamefont {I.~H.}\ \bibnamefont {Kim}},\ and\ \bibinfo
  {author} {\bibfnamefont {T.~H.}\ \bibnamefont {Hsieh}},\ }\href
  {https://doi.org/10.48550/arXiv.2206.13527} {\bibinfo {title} {Measurement as
  a shortcut to long-range entangled quantum matter}} (\bibinfo {year}
  {2022})\BibitemShut {NoStop}%
\bibitem [{\citenamefont {Tantivasadakarn}\ \emph {et~al.}(2023)\citenamefont
  {Tantivasadakarn}, \citenamefont {Vishwanath},\ and\ \citenamefont
  {Verresen}}]{hierarchy}%
  \BibitemOpen
  \bibfield  {author} {\bibinfo {author} {\bibfnamefont {N.}~\bibnamefont
  {Tantivasadakarn}}, \bibinfo {author} {\bibfnamefont {A.}~\bibnamefont
  {Vishwanath}},\ and\ \bibinfo {author} {\bibfnamefont {R.}~\bibnamefont
  {Verresen}},\ }\bibfield  {title} {\bibinfo {title} {Hierarchy of topological
  order from finite-depth unitaries, measurement, and feedforward},\ }\href
  {https://doi.org/10.1103/PRXQuantum.4.020339} {\bibfield  {journal} {\bibinfo
   {journal} {PRX Quantum}\ }\textbf {\bibinfo {volume} {4}},\ \bibinfo {pages}
  {020339} (\bibinfo {year} {2023})}\BibitemShut {NoStop}%
\bibitem [{\citenamefont {Piroli}\ \emph {et~al.}(2021)\citenamefont {Piroli},
  \citenamefont {Styliaris},\ and\ \citenamefont
  {Cirac}}]{piroli_quantum_2021}%
  \BibitemOpen
  \bibfield  {author} {\bibinfo {author} {\bibfnamefont {L.}~\bibnamefont
  {Piroli}}, \bibinfo {author} {\bibfnamefont {G.}~\bibnamefont {Styliaris}},\
  and\ \bibinfo {author} {\bibfnamefont {J.~I.}\ \bibnamefont {Cirac}},\
  }\bibfield  {title} {\bibinfo {title} {Quantum {Circuits} {Assisted} by
  {Local} {Operations} and {Classical} {Communication}: {Transformations} and
  {Phases} of {Matter}},\ }\href
  {https://doi.org/10.1103/PhysRevLett.127.220503} {\bibfield  {journal}
  {\bibinfo  {journal} {Physical Review Letters}\ }\textbf {\bibinfo {volume}
  {127}},\ \bibinfo {pages} {220503} (\bibinfo {year} {2021})}\BibitemShut
  {NoStop}%
\bibitem [{\citenamefont {Nigg}\ \emph {et~al.}(2014)\citenamefont {Nigg},
  \citenamefont {Mueller}, \citenamefont {Martinez}, \citenamefont {Schindler},
  \citenamefont {Hennrich}, \citenamefont {Monz}, \citenamefont
  {Martin-Delgado},\ and\ \citenamefont {Blatt}}]{nigg_experimental_2014}%
  \BibitemOpen
  \bibfield  {author} {\bibinfo {author} {\bibfnamefont {D.}~\bibnamefont
  {Nigg}}, \bibinfo {author} {\bibfnamefont {M.}~\bibnamefont {Mueller}},
  \bibinfo {author} {\bibfnamefont {E.~A.}\ \bibnamefont {Martinez}}, \bibinfo
  {author} {\bibfnamefont {P.}~\bibnamefont {Schindler}}, \bibinfo {author}
  {\bibfnamefont {M.}~\bibnamefont {Hennrich}}, \bibinfo {author}
  {\bibfnamefont {T.}~\bibnamefont {Monz}}, \bibinfo {author} {\bibfnamefont
  {M.~A.}\ \bibnamefont {Martin-Delgado}},\ and\ \bibinfo {author}
  {\bibfnamefont {R.}~\bibnamefont {Blatt}},\ }\bibfield  {title} {\bibinfo
  {title} {Experimental {Quantum} {Computations} on a {Topologically} {Encoded}
  {Qubit}},\ }\href {https://doi.org/10.1126/science.1253742} {\bibfield
  {journal} {\bibinfo  {journal} {Science}\ }\textbf {\bibinfo {volume}
  {345}},\ \bibinfo {pages} {302} (\bibinfo {year} {2014})}\BibitemShut
  {NoStop}%
\bibitem [{\citenamefont {Cramer}\ \emph {et~al.}(2016)\citenamefont {Cramer},
  \citenamefont {Kalb}, \citenamefont {Rol}, \citenamefont {Hensen},
  \citenamefont {Blok}, \citenamefont {Markham}, \citenamefont {Twitchen},
  \citenamefont {Hanson},\ and\ \citenamefont
  {Taminiau}}]{cramer_repeated_2016}%
  \BibitemOpen
  \bibfield  {author} {\bibinfo {author} {\bibfnamefont {J.}~\bibnamefont
  {Cramer}}, \bibinfo {author} {\bibfnamefont {N.}~\bibnamefont {Kalb}},
  \bibinfo {author} {\bibfnamefont {M.~A.}\ \bibnamefont {Rol}}, \bibinfo
  {author} {\bibfnamefont {B.}~\bibnamefont {Hensen}}, \bibinfo {author}
  {\bibfnamefont {M.~S.}\ \bibnamefont {Blok}}, \bibinfo {author}
  {\bibfnamefont {M.}~\bibnamefont {Markham}}, \bibinfo {author} {\bibfnamefont
  {D.~J.}\ \bibnamefont {Twitchen}}, \bibinfo {author} {\bibfnamefont
  {R.}~\bibnamefont {Hanson}},\ and\ \bibinfo {author} {\bibfnamefont {T.~H.}\
  \bibnamefont {Taminiau}},\ }\bibfield  {title} {\bibinfo {title} {Repeated
  quantum error correction on a continuously encoded qubit by real-time
  feedback},\ }\href {https://doi.org/10.1038/ncomms11526} {\bibfield
  {journal} {\bibinfo  {journal} {Nature Communications}\ }\textbf {\bibinfo
  {volume} {7}},\ \bibinfo {pages} {11526} (\bibinfo {year}
  {2016})}\BibitemShut {NoStop}%
\bibitem [{\citenamefont {Egan}\ \emph {et~al.}(2021)\citenamefont {Egan},
  \citenamefont {Debroy}, \citenamefont {Noel}, \citenamefont {Risinger},
  \citenamefont {Zhu}, \citenamefont {Biswas}, \citenamefont {Newman},
  \citenamefont {Li}, \citenamefont {Brown}, \citenamefont {Cetina},\ and\
  \citenamefont {Monroe}}]{egan_fault-tolerant_2021}%
  \BibitemOpen
  \bibfield  {author} {\bibinfo {author} {\bibfnamefont {L.}~\bibnamefont
  {Egan}}, \bibinfo {author} {\bibfnamefont {D.~M.}\ \bibnamefont {Debroy}},
  \bibinfo {author} {\bibfnamefont {C.}~\bibnamefont {Noel}}, \bibinfo {author}
  {\bibfnamefont {A.}~\bibnamefont {Risinger}}, \bibinfo {author}
  {\bibfnamefont {D.}~\bibnamefont {Zhu}}, \bibinfo {author} {\bibfnamefont
  {D.}~\bibnamefont {Biswas}}, \bibinfo {author} {\bibfnamefont
  {M.}~\bibnamefont {Newman}}, \bibinfo {author} {\bibfnamefont
  {M.}~\bibnamefont {Li}}, \bibinfo {author} {\bibfnamefont {K.~R.}\
  \bibnamefont {Brown}}, \bibinfo {author} {\bibfnamefont {M.}~\bibnamefont
  {Cetina}},\ and\ \bibinfo {author} {\bibfnamefont {C.}~\bibnamefont
  {Monroe}},\ }\href {https://doi.org/10.48550/arXiv.2009.11482} {\bibinfo
  {title} {Fault-{Tolerant} {Operation} of a {Quantum} {Error}-{Correction}
  {Code}}} (\bibinfo {year} {2021})\BibitemShut {NoStop}%
\bibitem [{\citenamefont {Chen}\ \emph {et~al.}(2021)\citenamefont {Chen},
  \citenamefont {Satzinger}, \citenamefont {Atalaya}, \citenamefont {Korotkov},
  \citenamefont {Dunsworth}, \citenamefont {Sank}, \citenamefont {Quintana},
  \citenamefont {McEwen}, \citenamefont {Barends}, \citenamefont {Klimov},
  \citenamefont {Hong}, \citenamefont {Jones}, \citenamefont {Petukhov},
  \citenamefont {Kafri}, \citenamefont {Demura}, \citenamefont {Burkett},
  \citenamefont {Gidney}, \citenamefont {Fowler}, \citenamefont {Putterman},
  \citenamefont {Aleiner}, \citenamefont {Arute}, \citenamefont {Arya},
  \citenamefont {Babbush}, \citenamefont {Bardin}, \citenamefont {Bengtsson},
  \citenamefont {Bourassa}, \citenamefont {Broughton}, \citenamefont {Buckley},
  \citenamefont {Buell}, \citenamefont {Bushnell}, \citenamefont {Chiaro},
  \citenamefont {Collins}, \citenamefont {Courtney}, \citenamefont {Derk},
  \citenamefont {Eppens}, \citenamefont {Erickson}, \citenamefont {Farhi},
  \citenamefont {Foxen}, \citenamefont {Giustina}, \citenamefont {Gross},
  \citenamefont {Harrigan}, \citenamefont {Harrington}, \citenamefont {Hilton},
  \citenamefont {Ho}, \citenamefont {Huang}, \citenamefont {Huggins},
  \citenamefont {Ioffe}, \citenamefont {Isakov}, \citenamefont {Jeffrey},
  \citenamefont {Jiang}, \citenamefont {Kechedzhi}, \citenamefont {Kim},
  \citenamefont {Kostritsa}, \citenamefont {Landhuis}, \citenamefont {Laptev},
  \citenamefont {Lucero}, \citenamefont {Martin}, \citenamefont {McClean},
  \citenamefont {McCourt}, \citenamefont {Mi}, \citenamefont {Miao},
  \citenamefont {Mohseni}, \citenamefont {Mruczkiewicz}, \citenamefont {Mutus},
  \citenamefont {Naaman}, \citenamefont {Neeley}, \citenamefont {Neill},
  \citenamefont {Newman}, \citenamefont {Niu}, \citenamefont {O'Brien},
  \citenamefont {Opremcak}, \citenamefont {Ostby}, \citenamefont {Pató},
  \citenamefont {Redd}, \citenamefont {Roushan}, \citenamefont {Rubin},
  \citenamefont {Shvarts}, \citenamefont {Strain}, \citenamefont {Szalay},
  \citenamefont {Trevithick}, \citenamefont {Villalonga}, \citenamefont
  {White}, \citenamefont {Yao}, \citenamefont {Yeh}, \citenamefont {Zalcman},
  \citenamefont {Neven}, \citenamefont {Boixo}, \citenamefont {Smelyanskiy},
  \citenamefont {Chen}, \citenamefont {Megrant},\ and\ \citenamefont
  {Kelly}}]{chen_exponential_2021}%
  \BibitemOpen
  \bibfield  {author} {\bibinfo {author} {\bibfnamefont {Z.}~\bibnamefont
  {Chen}}, \bibinfo {author} {\bibfnamefont {K.~J.}\ \bibnamefont {Satzinger}},
  \bibinfo {author} {\bibfnamefont {J.}~\bibnamefont {Atalaya}}, \bibinfo
  {author} {\bibfnamefont {A.~N.}\ \bibnamefont {Korotkov}}, \bibinfo {author}
  {\bibfnamefont {A.}~\bibnamefont {Dunsworth}}, \bibinfo {author}
  {\bibfnamefont {D.}~\bibnamefont {Sank}}, \bibinfo {author} {\bibfnamefont
  {C.}~\bibnamefont {Quintana}}, \bibinfo {author} {\bibfnamefont
  {M.}~\bibnamefont {McEwen}}, \bibinfo {author} {\bibfnamefont
  {R.}~\bibnamefont {Barends}}, \bibinfo {author} {\bibfnamefont {P.~V.}\
  \bibnamefont {Klimov}}, \bibinfo {author} {\bibfnamefont {S.}~\bibnamefont
  {Hong}}, \bibinfo {author} {\bibfnamefont {C.}~\bibnamefont {Jones}},
  \bibinfo {author} {\bibfnamefont {A.}~\bibnamefont {Petukhov}}, \bibinfo
  {author} {\bibfnamefont {D.}~\bibnamefont {Kafri}}, \bibinfo {author}
  {\bibfnamefont {S.}~\bibnamefont {Demura}}, \bibinfo {author} {\bibfnamefont
  {B.}~\bibnamefont {Burkett}}, \bibinfo {author} {\bibfnamefont
  {C.}~\bibnamefont {Gidney}}, \bibinfo {author} {\bibfnamefont {A.~G.}\
  \bibnamefont {Fowler}}, \bibinfo {author} {\bibfnamefont {H.}~\bibnamefont
  {Putterman}}, \bibinfo {author} {\bibfnamefont {I.}~\bibnamefont {Aleiner}},
  \bibinfo {author} {\bibfnamefont {F.}~\bibnamefont {Arute}}, \bibinfo
  {author} {\bibfnamefont {K.}~\bibnamefont {Arya}}, \bibinfo {author}
  {\bibfnamefont {R.}~\bibnamefont {Babbush}}, \bibinfo {author} {\bibfnamefont
  {J.~C.}\ \bibnamefont {Bardin}}, \bibinfo {author} {\bibfnamefont
  {A.}~\bibnamefont {Bengtsson}}, \bibinfo {author} {\bibfnamefont
  {A.}~\bibnamefont {Bourassa}}, \bibinfo {author} {\bibfnamefont
  {M.}~\bibnamefont {Broughton}}, \bibinfo {author} {\bibfnamefont {B.~B.}\
  \bibnamefont {Buckley}}, \bibinfo {author} {\bibfnamefont {D.~A.}\
  \bibnamefont {Buell}}, \bibinfo {author} {\bibfnamefont {N.}~\bibnamefont
  {Bushnell}}, \bibinfo {author} {\bibfnamefont {B.}~\bibnamefont {Chiaro}},
  \bibinfo {author} {\bibfnamefont {R.}~\bibnamefont {Collins}}, \bibinfo
  {author} {\bibfnamefont {W.}~\bibnamefont {Courtney}}, \bibinfo {author}
  {\bibfnamefont {A.~R.}\ \bibnamefont {Derk}}, \bibinfo {author}
  {\bibfnamefont {D.}~\bibnamefont {Eppens}}, \bibinfo {author} {\bibfnamefont
  {C.}~\bibnamefont {Erickson}}, \bibinfo {author} {\bibfnamefont
  {E.}~\bibnamefont {Farhi}}, \bibinfo {author} {\bibfnamefont
  {B.}~\bibnamefont {Foxen}}, \bibinfo {author} {\bibfnamefont
  {M.}~\bibnamefont {Giustina}}, \bibinfo {author} {\bibfnamefont {J.~A.}\
  \bibnamefont {Gross}}, \bibinfo {author} {\bibfnamefont {M.~P.}\ \bibnamefont
  {Harrigan}}, \bibinfo {author} {\bibfnamefont {S.~D.}\ \bibnamefont
  {Harrington}}, \bibinfo {author} {\bibfnamefont {J.}~\bibnamefont {Hilton}},
  \bibinfo {author} {\bibfnamefont {A.}~\bibnamefont {Ho}}, \bibinfo {author}
  {\bibfnamefont {T.}~\bibnamefont {Huang}}, \bibinfo {author} {\bibfnamefont
  {W.~J.}\ \bibnamefont {Huggins}}, \bibinfo {author} {\bibfnamefont {L.~B.}\
  \bibnamefont {Ioffe}}, \bibinfo {author} {\bibfnamefont {S.~V.}\ \bibnamefont
  {Isakov}}, \bibinfo {author} {\bibfnamefont {E.}~\bibnamefont {Jeffrey}},
  \bibinfo {author} {\bibfnamefont {Z.}~\bibnamefont {Jiang}}, \bibinfo
  {author} {\bibfnamefont {K.}~\bibnamefont {Kechedzhi}}, \bibinfo {author}
  {\bibfnamefont {S.}~\bibnamefont {Kim}}, \bibinfo {author} {\bibfnamefont
  {F.}~\bibnamefont {Kostritsa}}, \bibinfo {author} {\bibfnamefont
  {D.}~\bibnamefont {Landhuis}}, \bibinfo {author} {\bibfnamefont
  {P.}~\bibnamefont {Laptev}}, \bibinfo {author} {\bibfnamefont
  {E.}~\bibnamefont {Lucero}}, \bibinfo {author} {\bibfnamefont
  {O.}~\bibnamefont {Martin}}, \bibinfo {author} {\bibfnamefont {J.~R.}\
  \bibnamefont {McClean}}, \bibinfo {author} {\bibfnamefont {T.}~\bibnamefont
  {McCourt}}, \bibinfo {author} {\bibfnamefont {X.}~\bibnamefont {Mi}},
  \bibinfo {author} {\bibfnamefont {K.~C.}\ \bibnamefont {Miao}}, \bibinfo
  {author} {\bibfnamefont {M.}~\bibnamefont {Mohseni}}, \bibinfo {author}
  {\bibfnamefont {W.}~\bibnamefont {Mruczkiewicz}}, \bibinfo {author}
  {\bibfnamefont {J.}~\bibnamefont {Mutus}}, \bibinfo {author} {\bibfnamefont
  {O.}~\bibnamefont {Naaman}}, \bibinfo {author} {\bibfnamefont
  {M.}~\bibnamefont {Neeley}}, \bibinfo {author} {\bibfnamefont
  {C.}~\bibnamefont {Neill}}, \bibinfo {author} {\bibfnamefont
  {M.}~\bibnamefont {Newman}}, \bibinfo {author} {\bibfnamefont {M.~Y.}\
  \bibnamefont {Niu}}, \bibinfo {author} {\bibfnamefont {T.~E.}\ \bibnamefont
  {O'Brien}}, \bibinfo {author} {\bibfnamefont {A.}~\bibnamefont {Opremcak}},
  \bibinfo {author} {\bibfnamefont {E.}~\bibnamefont {Ostby}}, \bibinfo
  {author} {\bibfnamefont {B.}~\bibnamefont {Pató}}, \bibinfo {author}
  {\bibfnamefont {N.}~\bibnamefont {Redd}}, \bibinfo {author} {\bibfnamefont
  {P.}~\bibnamefont {Roushan}}, \bibinfo {author} {\bibfnamefont {N.~C.}\
  \bibnamefont {Rubin}}, \bibinfo {author} {\bibfnamefont {V.}~\bibnamefont
  {Shvarts}}, \bibinfo {author} {\bibfnamefont {D.}~\bibnamefont {Strain}},
  \bibinfo {author} {\bibfnamefont {M.}~\bibnamefont {Szalay}}, \bibinfo
  {author} {\bibfnamefont {M.~D.}\ \bibnamefont {Trevithick}}, \bibinfo
  {author} {\bibfnamefont {B.}~\bibnamefont {Villalonga}}, \bibinfo {author}
  {\bibfnamefont {T.}~\bibnamefont {White}}, \bibinfo {author} {\bibfnamefont
  {Z.~J.}\ \bibnamefont {Yao}}, \bibinfo {author} {\bibfnamefont
  {P.}~\bibnamefont {Yeh}}, \bibinfo {author} {\bibfnamefont {A.}~\bibnamefont
  {Zalcman}}, \bibinfo {author} {\bibfnamefont {H.}~\bibnamefont {Neven}},
  \bibinfo {author} {\bibfnamefont {S.}~\bibnamefont {Boixo}}, \bibinfo
  {author} {\bibfnamefont {V.}~\bibnamefont {Smelyanskiy}}, \bibinfo {author}
  {\bibfnamefont {Y.}~\bibnamefont {Chen}}, \bibinfo {author} {\bibfnamefont
  {A.}~\bibnamefont {Megrant}},\ and\ \bibinfo {author} {\bibfnamefont
  {J.}~\bibnamefont {Kelly}},\ }\bibfield  {title} {\bibinfo {title}
  {Exponential suppression of bit or phase flip errors with repetitive error
  correction},\ }\href {https://doi.org/10.1038/s41586-021-03588-y} {\bibfield
  {journal} {\bibinfo  {journal} {Nature}\ }\textbf {\bibinfo {volume} {595}},\
  \bibinfo {pages} {383} (\bibinfo {year} {2021})}\BibitemShut {NoStop}%
\bibitem [{\citenamefont {Ryan-Anderson}\ \emph {et~al.}(2021)\citenamefont
  {Ryan-Anderson}, \citenamefont {Bohnet}, \citenamefont {Lee}, \citenamefont
  {Gresh}, \citenamefont {Hankin}, \citenamefont {Gaebler}, \citenamefont
  {Francois}, \citenamefont {Chernoguzov}, \citenamefont {Lucchetti},
  \citenamefont {Brown}, \citenamefont {Gatterman}, \citenamefont {Halit},
  \citenamefont {Gilmore}, \citenamefont {Gerber}, \citenamefont {Neyenhuis},
  \citenamefont {Hayes},\ and\ \citenamefont
  {Stutz}}]{ryan-anderson_realization_2021}%
  \BibitemOpen
  \bibfield  {author} {\bibinfo {author} {\bibfnamefont {C.}~\bibnamefont
  {Ryan-Anderson}}, \bibinfo {author} {\bibfnamefont {J.~G.}\ \bibnamefont
  {Bohnet}}, \bibinfo {author} {\bibfnamefont {K.}~\bibnamefont {Lee}},
  \bibinfo {author} {\bibfnamefont {D.}~\bibnamefont {Gresh}}, \bibinfo
  {author} {\bibfnamefont {A.}~\bibnamefont {Hankin}}, \bibinfo {author}
  {\bibfnamefont {J.~P.}\ \bibnamefont {Gaebler}}, \bibinfo {author}
  {\bibfnamefont {D.}~\bibnamefont {Francois}}, \bibinfo {author}
  {\bibfnamefont {A.}~\bibnamefont {Chernoguzov}}, \bibinfo {author}
  {\bibfnamefont {D.}~\bibnamefont {Lucchetti}}, \bibinfo {author}
  {\bibfnamefont {N.~C.}\ \bibnamefont {Brown}}, \bibinfo {author}
  {\bibfnamefont {T.~M.}\ \bibnamefont {Gatterman}}, \bibinfo {author}
  {\bibfnamefont {S.~K.}\ \bibnamefont {Halit}}, \bibinfo {author}
  {\bibfnamefont {K.}~\bibnamefont {Gilmore}}, \bibinfo {author} {\bibfnamefont
  {J.}~\bibnamefont {Gerber}}, \bibinfo {author} {\bibfnamefont
  {B.}~\bibnamefont {Neyenhuis}}, \bibinfo {author} {\bibfnamefont
  {D.}~\bibnamefont {Hayes}},\ and\ \bibinfo {author} {\bibfnamefont {R.~P.}\
  \bibnamefont {Stutz}},\ }\href {https://doi.org/10.48550/arXiv.2107.07505}
  {\bibinfo {title} {Realization of real-time fault-tolerant quantum error
  correction}} (\bibinfo {year} {2021})\BibitemShut {NoStop}%
\bibitem [{\citenamefont {Ryan-Anderson}\ \emph {et~al.}(2022)\citenamefont
  {Ryan-Anderson}, \citenamefont {Brown}, \citenamefont {Allman}, \citenamefont
  {Arkin}, \citenamefont {Asa-Attuah}, \citenamefont {Baldwin}, \citenamefont
  {Berg}, \citenamefont {Bohnet}, \citenamefont {Braxton}, \citenamefont
  {Burdick}, \citenamefont {Campora}, \citenamefont {Chernoguzov},
  \citenamefont {Esposito}, \citenamefont {Evans}, \citenamefont {Francois},
  \citenamefont {Gaebler}, \citenamefont {Gatterman}, \citenamefont {Gerber},
  \citenamefont {Gilmore}, \citenamefont {Gresh}, \citenamefont {Hall},
  \citenamefont {Hankin}, \citenamefont {Hostetter}, \citenamefont {Lucchetti},
  \citenamefont {Mayer}, \citenamefont {Myers}, \citenamefont {Neyenhuis},
  \citenamefont {Santiago}, \citenamefont {Sedlacek}, \citenamefont {Skripka},
  \citenamefont {Slattery}, \citenamefont {Stutz}, \citenamefont {Tait},
  \citenamefont {Tobey}, \citenamefont {Vittorini}, \citenamefont {Walker},\
  and\ \citenamefont {Hayes}}]{ryan-anderson_implementing_2022}%
  \BibitemOpen
  \bibfield  {author} {\bibinfo {author} {\bibfnamefont {C.}~\bibnamefont
  {Ryan-Anderson}}, \bibinfo {author} {\bibfnamefont {N.~C.}\ \bibnamefont
  {Brown}}, \bibinfo {author} {\bibfnamefont {M.~S.}\ \bibnamefont {Allman}},
  \bibinfo {author} {\bibfnamefont {B.}~\bibnamefont {Arkin}}, \bibinfo
  {author} {\bibfnamefont {G.}~\bibnamefont {Asa-Attuah}}, \bibinfo {author}
  {\bibfnamefont {C.}~\bibnamefont {Baldwin}}, \bibinfo {author} {\bibfnamefont
  {J.}~\bibnamefont {Berg}}, \bibinfo {author} {\bibfnamefont {J.~G.}\
  \bibnamefont {Bohnet}}, \bibinfo {author} {\bibfnamefont {S.}~\bibnamefont
  {Braxton}}, \bibinfo {author} {\bibfnamefont {N.}~\bibnamefont {Burdick}},
  \bibinfo {author} {\bibfnamefont {J.~P.}\ \bibnamefont {Campora}}, \bibinfo
  {author} {\bibfnamefont {A.}~\bibnamefont {Chernoguzov}}, \bibinfo {author}
  {\bibfnamefont {J.}~\bibnamefont {Esposito}}, \bibinfo {author}
  {\bibfnamefont {B.}~\bibnamefont {Evans}}, \bibinfo {author} {\bibfnamefont
  {D.}~\bibnamefont {Francois}}, \bibinfo {author} {\bibfnamefont {J.~P.}\
  \bibnamefont {Gaebler}}, \bibinfo {author} {\bibfnamefont {T.~M.}\
  \bibnamefont {Gatterman}}, \bibinfo {author} {\bibfnamefont {J.}~\bibnamefont
  {Gerber}}, \bibinfo {author} {\bibfnamefont {K.}~\bibnamefont {Gilmore}},
  \bibinfo {author} {\bibfnamefont {D.}~\bibnamefont {Gresh}}, \bibinfo
  {author} {\bibfnamefont {A.}~\bibnamefont {Hall}}, \bibinfo {author}
  {\bibfnamefont {A.}~\bibnamefont {Hankin}}, \bibinfo {author} {\bibfnamefont
  {J.}~\bibnamefont {Hostetter}}, \bibinfo {author} {\bibfnamefont
  {D.}~\bibnamefont {Lucchetti}}, \bibinfo {author} {\bibfnamefont
  {K.}~\bibnamefont {Mayer}}, \bibinfo {author} {\bibfnamefont
  {J.}~\bibnamefont {Myers}}, \bibinfo {author} {\bibfnamefont
  {B.}~\bibnamefont {Neyenhuis}}, \bibinfo {author} {\bibfnamefont
  {J.}~\bibnamefont {Santiago}}, \bibinfo {author} {\bibfnamefont
  {J.}~\bibnamefont {Sedlacek}}, \bibinfo {author} {\bibfnamefont
  {T.}~\bibnamefont {Skripka}}, \bibinfo {author} {\bibfnamefont
  {A.}~\bibnamefont {Slattery}}, \bibinfo {author} {\bibfnamefont {R.~P.}\
  \bibnamefont {Stutz}}, \bibinfo {author} {\bibfnamefont {J.}~\bibnamefont
  {Tait}}, \bibinfo {author} {\bibfnamefont {R.}~\bibnamefont {Tobey}},
  \bibinfo {author} {\bibfnamefont {G.}~\bibnamefont {Vittorini}}, \bibinfo
  {author} {\bibfnamefont {J.}~\bibnamefont {Walker}},\ and\ \bibinfo {author}
  {\bibfnamefont {D.}~\bibnamefont {Hayes}},\ }\href
  {https://doi.org/10.48550/arXiv.2208.01863} {\bibinfo {title} {Implementing
  {Fault}-tolerant {Entangling} {Gates} on the {Five}-qubit {Code} and the
  {Color} {Code}}} (\bibinfo {year} {2022})\BibitemShut {NoStop}%
\bibitem [{\citenamefont {Aguado}\ \emph {et~al.}(2008)\citenamefont {Aguado},
  \citenamefont {Brennen}, \citenamefont {Verstraete},\ and\ \citenamefont
  {Cirac}}]{aguado_creation_2008}%
  \BibitemOpen
  \bibfield  {author} {\bibinfo {author} {\bibfnamefont {M.}~\bibnamefont
  {Aguado}}, \bibinfo {author} {\bibfnamefont {G.~K.}\ \bibnamefont {Brennen}},
  \bibinfo {author} {\bibfnamefont {F.}~\bibnamefont {Verstraete}},\ and\
  \bibinfo {author} {\bibfnamefont {J.~I.}\ \bibnamefont {Cirac}},\ }\bibfield
  {title} {\bibinfo {title} {Creation, manipulation, and detection of {Abelian}
  and non-{Abelian} anyons in optical lattices},\ }\href
  {https://doi.org/10.1103/PhysRevLett.101.260501} {\bibfield  {journal}
  {\bibinfo  {journal} {Physical Review Letters}\ }\textbf {\bibinfo {volume}
  {101}},\ \bibinfo {pages} {260501} (\bibinfo {year} {2008})}\BibitemShut
  {NoStop}%
\bibitem [{\citenamefont {Satzinger}\ \emph {et~al.}(2021)\citenamefont
  {Satzinger}, \citenamefont {Liu}, \citenamefont {Smith}, \citenamefont
  {Knapp}, \citenamefont {Newman}, \citenamefont {Jones}, \citenamefont {Chen},
  \citenamefont {Quintana}, \citenamefont {Mi}, \citenamefont {Dunsworth},
  \citenamefont {Gidney}, \citenamefont {Aleiner}, \citenamefont {Arute},
  \citenamefont {Arya}, \citenamefont {Atalaya}, \citenamefont {Babbush},
  \citenamefont {Bardin}, \citenamefont {Barends}, \citenamefont {Basso},
  \citenamefont {Bengtsson}, \citenamefont {Bilmes}, \citenamefont {Broughton},
  \citenamefont {Buckley}, \citenamefont {Buell}, \citenamefont {Burkett},
  \citenamefont {Bushnell}, \citenamefont {Chiaro}, \citenamefont {Collins},
  \citenamefont {Courtney}, \citenamefont {Demura}, \citenamefont {Derk},
  \citenamefont {Eppens}, \citenamefont {Erickson}, \citenamefont {Farhi},
  \citenamefont {Foaro}, \citenamefont {Fowler}, \citenamefont {Foxen},
  \citenamefont {Giustina}, \citenamefont {Greene}, \citenamefont {Gross},
  \citenamefont {Harrigan}, \citenamefont {Harrington}, \citenamefont {Hilton},
  \citenamefont {Hong}, \citenamefont {Huang}, \citenamefont {Huggins},
  \citenamefont {Ioffe}, \citenamefont {Isakov}, \citenamefont {Jeffrey},
  \citenamefont {Jiang}, \citenamefont {Kafri}, \citenamefont {Kechedzhi},
  \citenamefont {Khattar}, \citenamefont {Kim}, \citenamefont {Klimov},
  \citenamefont {Korotkov}, \citenamefont {Kostritsa}, \citenamefont
  {Landhuis}, \citenamefont {Laptev}, \citenamefont {Locharla}, \citenamefont
  {Lucero}, \citenamefont {Martin}, \citenamefont {McClean}, \citenamefont
  {McEwen}, \citenamefont {Miao}, \citenamefont {Mohseni}, \citenamefont
  {Montazeri}, \citenamefont {Mruczkiewicz}, \citenamefont {Mutus},
  \citenamefont {Naaman}, \citenamefont {Neeley}, \citenamefont {Neill},
  \citenamefont {Niu}, \citenamefont {O'Brien}, \citenamefont {Opremcak},
  \citenamefont {Pató}, \citenamefont {Petukhov}, \citenamefont {Rubin},
  \citenamefont {Sank}, \citenamefont {Shvarts}, \citenamefont {Strain},
  \citenamefont {Szalay}, \citenamefont {Villalonga}, \citenamefont {White},
  \citenamefont {Yao}, \citenamefont {Yeh}, \citenamefont {Yoo}, \citenamefont
  {Zalcman}, \citenamefont {Neven}, \citenamefont {Boixo}, \citenamefont
  {Megrant}, \citenamefont {Chen}, \citenamefont {Kelly}, \citenamefont
  {Smelyanskiy}, \citenamefont {Kitaev}, \citenamefont {Knap}, \citenamefont
  {Pollmann},\ and\ \citenamefont {Roushan}}]{satzinger_realizing_2021}%
  \BibitemOpen
  \bibfield  {author} {\bibinfo {author} {\bibfnamefont {K.~J.}\ \bibnamefont
  {Satzinger}}, \bibinfo {author} {\bibfnamefont {Y.}~\bibnamefont {Liu}},
  \bibinfo {author} {\bibfnamefont {A.}~\bibnamefont {Smith}}, \bibinfo
  {author} {\bibfnamefont {C.}~\bibnamefont {Knapp}}, \bibinfo {author}
  {\bibfnamefont {M.}~\bibnamefont {Newman}}, \bibinfo {author} {\bibfnamefont
  {C.}~\bibnamefont {Jones}}, \bibinfo {author} {\bibfnamefont
  {Z.}~\bibnamefont {Chen}}, \bibinfo {author} {\bibfnamefont {C.}~\bibnamefont
  {Quintana}}, \bibinfo {author} {\bibfnamefont {X.}~\bibnamefont {Mi}},
  \bibinfo {author} {\bibfnamefont {A.}~\bibnamefont {Dunsworth}}, \bibinfo
  {author} {\bibfnamefont {C.}~\bibnamefont {Gidney}}, \bibinfo {author}
  {\bibfnamefont {I.}~\bibnamefont {Aleiner}}, \bibinfo {author} {\bibfnamefont
  {F.}~\bibnamefont {Arute}}, \bibinfo {author} {\bibfnamefont
  {K.}~\bibnamefont {Arya}}, \bibinfo {author} {\bibfnamefont {J.}~\bibnamefont
  {Atalaya}}, \bibinfo {author} {\bibfnamefont {R.}~\bibnamefont {Babbush}},
  \bibinfo {author} {\bibfnamefont {J.~C.}\ \bibnamefont {Bardin}}, \bibinfo
  {author} {\bibfnamefont {R.}~\bibnamefont {Barends}}, \bibinfo {author}
  {\bibfnamefont {J.}~\bibnamefont {Basso}}, \bibinfo {author} {\bibfnamefont
  {A.}~\bibnamefont {Bengtsson}}, \bibinfo {author} {\bibfnamefont
  {A.}~\bibnamefont {Bilmes}}, \bibinfo {author} {\bibfnamefont
  {M.}~\bibnamefont {Broughton}}, \bibinfo {author} {\bibfnamefont {B.~B.}\
  \bibnamefont {Buckley}}, \bibinfo {author} {\bibfnamefont {D.~A.}\
  \bibnamefont {Buell}}, \bibinfo {author} {\bibfnamefont {B.}~\bibnamefont
  {Burkett}}, \bibinfo {author} {\bibfnamefont {N.}~\bibnamefont {Bushnell}},
  \bibinfo {author} {\bibfnamefont {B.}~\bibnamefont {Chiaro}}, \bibinfo
  {author} {\bibfnamefont {R.}~\bibnamefont {Collins}}, \bibinfo {author}
  {\bibfnamefont {W.}~\bibnamefont {Courtney}}, \bibinfo {author}
  {\bibfnamefont {S.}~\bibnamefont {Demura}}, \bibinfo {author} {\bibfnamefont
  {A.~R.}\ \bibnamefont {Derk}}, \bibinfo {author} {\bibfnamefont
  {D.}~\bibnamefont {Eppens}}, \bibinfo {author} {\bibfnamefont
  {C.}~\bibnamefont {Erickson}}, \bibinfo {author} {\bibfnamefont
  {E.}~\bibnamefont {Farhi}}, \bibinfo {author} {\bibfnamefont
  {L.}~\bibnamefont {Foaro}}, \bibinfo {author} {\bibfnamefont {A.~G.}\
  \bibnamefont {Fowler}}, \bibinfo {author} {\bibfnamefont {B.}~\bibnamefont
  {Foxen}}, \bibinfo {author} {\bibfnamefont {M.}~\bibnamefont {Giustina}},
  \bibinfo {author} {\bibfnamefont {A.}~\bibnamefont {Greene}}, \bibinfo
  {author} {\bibfnamefont {J.~A.}\ \bibnamefont {Gross}}, \bibinfo {author}
  {\bibfnamefont {M.~P.}\ \bibnamefont {Harrigan}}, \bibinfo {author}
  {\bibfnamefont {S.~D.}\ \bibnamefont {Harrington}}, \bibinfo {author}
  {\bibfnamefont {J.}~\bibnamefont {Hilton}}, \bibinfo {author} {\bibfnamefont
  {S.}~\bibnamefont {Hong}}, \bibinfo {author} {\bibfnamefont {T.}~\bibnamefont
  {Huang}}, \bibinfo {author} {\bibfnamefont {W.~J.}\ \bibnamefont {Huggins}},
  \bibinfo {author} {\bibfnamefont {L.~B.}\ \bibnamefont {Ioffe}}, \bibinfo
  {author} {\bibfnamefont {S.~V.}\ \bibnamefont {Isakov}}, \bibinfo {author}
  {\bibfnamefont {E.}~\bibnamefont {Jeffrey}}, \bibinfo {author} {\bibfnamefont
  {Z.}~\bibnamefont {Jiang}}, \bibinfo {author} {\bibfnamefont
  {D.}~\bibnamefont {Kafri}}, \bibinfo {author} {\bibfnamefont
  {K.}~\bibnamefont {Kechedzhi}}, \bibinfo {author} {\bibfnamefont
  {T.}~\bibnamefont {Khattar}}, \bibinfo {author} {\bibfnamefont
  {S.}~\bibnamefont {Kim}}, \bibinfo {author} {\bibfnamefont {P.~V.}\
  \bibnamefont {Klimov}}, \bibinfo {author} {\bibfnamefont {A.~N.}\
  \bibnamefont {Korotkov}}, \bibinfo {author} {\bibfnamefont {F.}~\bibnamefont
  {Kostritsa}}, \bibinfo {author} {\bibfnamefont {D.}~\bibnamefont {Landhuis}},
  \bibinfo {author} {\bibfnamefont {P.}~\bibnamefont {Laptev}}, \bibinfo
  {author} {\bibfnamefont {A.}~\bibnamefont {Locharla}}, \bibinfo {author}
  {\bibfnamefont {E.}~\bibnamefont {Lucero}}, \bibinfo {author} {\bibfnamefont
  {O.}~\bibnamefont {Martin}}, \bibinfo {author} {\bibfnamefont {J.~R.}\
  \bibnamefont {McClean}}, \bibinfo {author} {\bibfnamefont {M.}~\bibnamefont
  {McEwen}}, \bibinfo {author} {\bibfnamefont {K.~C.}\ \bibnamefont {Miao}},
  \bibinfo {author} {\bibfnamefont {M.}~\bibnamefont {Mohseni}}, \bibinfo
  {author} {\bibfnamefont {S.}~\bibnamefont {Montazeri}}, \bibinfo {author}
  {\bibfnamefont {W.}~\bibnamefont {Mruczkiewicz}}, \bibinfo {author}
  {\bibfnamefont {J.}~\bibnamefont {Mutus}}, \bibinfo {author} {\bibfnamefont
  {O.}~\bibnamefont {Naaman}}, \bibinfo {author} {\bibfnamefont
  {M.}~\bibnamefont {Neeley}}, \bibinfo {author} {\bibfnamefont
  {C.}~\bibnamefont {Neill}}, \bibinfo {author} {\bibfnamefont {M.~Y.}\
  \bibnamefont {Niu}}, \bibinfo {author} {\bibfnamefont {T.~E.}\ \bibnamefont
  {O'Brien}}, \bibinfo {author} {\bibfnamefont {A.}~\bibnamefont {Opremcak}},
  \bibinfo {author} {\bibfnamefont {B.}~\bibnamefont {Pató}}, \bibinfo
  {author} {\bibfnamefont {A.}~\bibnamefont {Petukhov}}, \bibinfo {author}
  {\bibfnamefont {N.~C.}\ \bibnamefont {Rubin}}, \bibinfo {author}
  {\bibfnamefont {D.}~\bibnamefont {Sank}}, \bibinfo {author} {\bibfnamefont
  {V.}~\bibnamefont {Shvarts}}, \bibinfo {author} {\bibfnamefont
  {D.}~\bibnamefont {Strain}}, \bibinfo {author} {\bibfnamefont
  {M.}~\bibnamefont {Szalay}}, \bibinfo {author} {\bibfnamefont
  {B.}~\bibnamefont {Villalonga}}, \bibinfo {author} {\bibfnamefont {T.~C.}\
  \bibnamefont {White}}, \bibinfo {author} {\bibfnamefont {Z.}~\bibnamefont
  {Yao}}, \bibinfo {author} {\bibfnamefont {P.}~\bibnamefont {Yeh}}, \bibinfo
  {author} {\bibfnamefont {J.}~\bibnamefont {Yoo}}, \bibinfo {author}
  {\bibfnamefont {A.}~\bibnamefont {Zalcman}}, \bibinfo {author} {\bibfnamefont
  {H.}~\bibnamefont {Neven}}, \bibinfo {author} {\bibfnamefont
  {S.}~\bibnamefont {Boixo}}, \bibinfo {author} {\bibfnamefont
  {A.}~\bibnamefont {Megrant}}, \bibinfo {author} {\bibfnamefont
  {Y.}~\bibnamefont {Chen}}, \bibinfo {author} {\bibfnamefont {J.}~\bibnamefont
  {Kelly}}, \bibinfo {author} {\bibfnamefont {V.}~\bibnamefont {Smelyanskiy}},
  \bibinfo {author} {\bibfnamefont {A.}~\bibnamefont {Kitaev}}, \bibinfo
  {author} {\bibfnamefont {M.}~\bibnamefont {Knap}}, \bibinfo {author}
  {\bibfnamefont {F.}~\bibnamefont {Pollmann}},\ and\ \bibinfo {author}
  {\bibfnamefont {P.}~\bibnamefont {Roushan}},\ }\bibfield  {title} {\bibinfo
  {title} {Realizing topologically ordered states on a quantum processor},\
  }\href {https://doi.org/10.1126/science.abi8378} {\bibfield  {journal}
  {\bibinfo  {journal} {Science}\ }\textbf {\bibinfo {volume} {374}},\ \bibinfo
  {pages} {1237} (\bibinfo {year} {2021})}\BibitemShut {NoStop}%
\bibitem [{\citenamefont {Bluvstein}\ \emph {et~al.}(2022)\citenamefont
  {Bluvstein}, \citenamefont {Levine}, \citenamefont {Semeghini}, \citenamefont
  {Wang}, \citenamefont {Ebadi}, \citenamefont {Kalinowski}, \citenamefont
  {Keesling}, \citenamefont {Maskara}, \citenamefont {Pichler}, \citenamefont
  {Greiner}, \citenamefont {Vuletic},\ and\ \citenamefont
  {Lukin}}]{bluvstein_quantum_2022}%
  \BibitemOpen
  \bibfield  {author} {\bibinfo {author} {\bibfnamefont {D.}~\bibnamefont
  {Bluvstein}}, \bibinfo {author} {\bibfnamefont {H.}~\bibnamefont {Levine}},
  \bibinfo {author} {\bibfnamefont {G.}~\bibnamefont {Semeghini}}, \bibinfo
  {author} {\bibfnamefont {T.~T.}\ \bibnamefont {Wang}}, \bibinfo {author}
  {\bibfnamefont {S.}~\bibnamefont {Ebadi}}, \bibinfo {author} {\bibfnamefont
  {M.}~\bibnamefont {Kalinowski}}, \bibinfo {author} {\bibfnamefont
  {A.}~\bibnamefont {Keesling}}, \bibinfo {author} {\bibfnamefont
  {N.}~\bibnamefont {Maskara}}, \bibinfo {author} {\bibfnamefont
  {H.}~\bibnamefont {Pichler}}, \bibinfo {author} {\bibfnamefont
  {M.}~\bibnamefont {Greiner}}, \bibinfo {author} {\bibfnamefont
  {V.}~\bibnamefont {Vuletic}},\ and\ \bibinfo {author} {\bibfnamefont {M.~D.}\
  \bibnamefont {Lukin}},\ }\bibfield  {title} {\bibinfo {title} {A quantum
  processor based on coherent transport of entangled atom arrays},\ }\href
  {https://doi.org/10.1038/s41586-022-04592-6} {\bibfield  {journal} {\bibinfo
  {journal} {Nature}\ }\textbf {\bibinfo {volume} {604}},\ \bibinfo {pages}
  {451} (\bibinfo {year} {2022})}\BibitemShut {NoStop}%
\bibitem [{\citenamefont {Andersen}\ \emph {et~al.}(2022)\citenamefont
  {Andersen}, \citenamefont {Lensky}, \citenamefont {Kechedzhi}, \citenamefont
  {Drozdov}, \citenamefont {Bengtsson}, \citenamefont {Hong}, \citenamefont
  {Morvan}, \citenamefont {Mi}, \citenamefont {Opremcak}, \citenamefont
  {Acharya}, \citenamefont {Allen}, \citenamefont {Ansmann}, \citenamefont
  {Arute}, \citenamefont {Arya}, \citenamefont {Asfaw}, \citenamefont
  {Atalaya}, \citenamefont {Babbush}, \citenamefont {Bacon}, \citenamefont
  {Bardin}, \citenamefont {Bortoli}, \citenamefont {Bourassa}, \citenamefont
  {Bovaird}, \citenamefont {Brill}, \citenamefont {Broughton}, \citenamefont
  {Buckley}, \citenamefont {Buell}, \citenamefont {Burger}, \citenamefont
  {Burkett}, \citenamefont {Bushnell}, \citenamefont {Chen}, \citenamefont
  {Chiaro}, \citenamefont {Chik}, \citenamefont {Chou}, \citenamefont {Cogan},
  \citenamefont {Collins}, \citenamefont {Conner}, \citenamefont {Courtney},
  \citenamefont {Crook}, \citenamefont {Curtin}, \citenamefont {Debroy},
  \citenamefont {Barba}, \citenamefont {Demura}, \citenamefont {Dunsworth},
  \citenamefont {Eppens}, \citenamefont {Erickson}, \citenamefont {Faoro},
  \citenamefont {Farhi}, \citenamefont {Fatemi}, \citenamefont {Ferreira},
  \citenamefont {Burgos}, \citenamefont {Forati}, \citenamefont {Fowler},
  \citenamefont {Foxen}, \citenamefont {Giang}, \citenamefont {Gidney},
  \citenamefont {Gilboa}, \citenamefont {Giustina}, \citenamefont {Gosula},
  \citenamefont {Dau}, \citenamefont {Gross}, \citenamefont {Habegger},
  \citenamefont {Hamilton}, \citenamefont {Hansen}, \citenamefont {Harrigan},
  \citenamefont {Harrington}, \citenamefont {Heu}, \citenamefont {Hilton},
  \citenamefont {Hoffmann}, \citenamefont {Huang}, \citenamefont {Huff},
  \citenamefont {Huggins}, \citenamefont {Ioffe}, \citenamefont {Isakov},
  \citenamefont {Iveland}, \citenamefont {Jeffrey}, \citenamefont {Jiang},
  \citenamefont {Jones}, \citenamefont {Juhas}, \citenamefont {Kafri},
  \citenamefont {Khattar}, \citenamefont {Khezri}, \citenamefont {Kieferová},
  \citenamefont {Kim}, \citenamefont {Kitaev}, \citenamefont {Klimov},
  \citenamefont {Klots}, \citenamefont {Korotkov}, \citenamefont {Kostritsa},
  \citenamefont {Kreikebaum}, \citenamefont {Landhuis}, \citenamefont {Laptev},
  \citenamefont {Lau}, \citenamefont {Laws}, \citenamefont {Lee}, \citenamefont
  {Lee}, \citenamefont {Lester}, \citenamefont {Lill}, \citenamefont {Liu},
  \citenamefont {Locharla}, \citenamefont {Lucero}, \citenamefont {Malone},
  \citenamefont {Martin}, \citenamefont {McClean}, \citenamefont {McCourt},
  \citenamefont {McEwen}, \citenamefont {Miao}, \citenamefont {Mieszala},
  \citenamefont {Mohseni}, \citenamefont {Montazeri}, \citenamefont {Mount},
  \citenamefont {Movassagh}, \citenamefont {Mruczkiewicz}, \citenamefont
  {Naaman}, \citenamefont {Neeley}, \citenamefont {Neill}, \citenamefont
  {Nersisyan}, \citenamefont {Newman}, \citenamefont {Ng}, \citenamefont
  {Nguyen}, \citenamefont {Nguyen}, \citenamefont {Niu}, \citenamefont
  {O'Brien}, \citenamefont {Omonije}, \citenamefont {Petukhov}, \citenamefont
  {Potter}, \citenamefont {Pryadko}, \citenamefont {Quintana}, \citenamefont
  {Rocque}, \citenamefont {Rubin}, \citenamefont {Saei}, \citenamefont {Sank},
  \citenamefont {Sankaragomathi}, \citenamefont {Satzinger}, \citenamefont
  {Schurkus}, \citenamefont {Schuster}, \citenamefont {Shearn}, \citenamefont
  {Shorter}, \citenamefont {Shutty}, \citenamefont {Shvarts}, \citenamefont
  {Skruzny}, \citenamefont {Smith}, \citenamefont {Somma}, \citenamefont
  {Sterling}, \citenamefont {Strain}, \citenamefont {Szalay}, \citenamefont
  {Torres}, \citenamefont {Vidal}, \citenamefont {Villalonga}, \citenamefont
  {Heidweiller}, \citenamefont {White}, \citenamefont {Woo}, \citenamefont
  {Xing}, \citenamefont {Yao}, \citenamefont {Yeh}, \citenamefont {Yoo},
  \citenamefont {Young}, \citenamefont {Zalcman}, \citenamefont {Zhang},
  \citenamefont {Zhu}, \citenamefont {Zobrist}, \citenamefont {Neven},
  \citenamefont {Boixo}, \citenamefont {Megrant}, \citenamefont {Kelly},
  \citenamefont {Chen}, \citenamefont {Smelyanskiy}, \citenamefont {Kim},
  \citenamefont {Aleiner},\ and\ \citenamefont
  {Roushan}}]{andersen_observation_2022}%
  \BibitemOpen
  \bibfield  {author} {\bibinfo {author} {\bibfnamefont {T.~I.}\ \bibnamefont
  {Andersen}}, \bibinfo {author} {\bibfnamefont {Y.~D.}\ \bibnamefont
  {Lensky}}, \bibinfo {author} {\bibfnamefont {K.}~\bibnamefont {Kechedzhi}},
  \bibinfo {author} {\bibfnamefont {I.}~\bibnamefont {Drozdov}}, \bibinfo
  {author} {\bibfnamefont {A.}~\bibnamefont {Bengtsson}}, \bibinfo {author}
  {\bibfnamefont {S.}~\bibnamefont {Hong}}, \bibinfo {author} {\bibfnamefont
  {A.}~\bibnamefont {Morvan}}, \bibinfo {author} {\bibfnamefont
  {X.}~\bibnamefont {Mi}}, \bibinfo {author} {\bibfnamefont {A.}~\bibnamefont
  {Opremcak}}, \bibinfo {author} {\bibfnamefont {R.}~\bibnamefont {Acharya}},
  \bibinfo {author} {\bibfnamefont {R.}~\bibnamefont {Allen}}, \bibinfo
  {author} {\bibfnamefont {M.}~\bibnamefont {Ansmann}}, \bibinfo {author}
  {\bibfnamefont {F.}~\bibnamefont {Arute}}, \bibinfo {author} {\bibfnamefont
  {K.}~\bibnamefont {Arya}}, \bibinfo {author} {\bibfnamefont {A.}~\bibnamefont
  {Asfaw}}, \bibinfo {author} {\bibfnamefont {J.}~\bibnamefont {Atalaya}},
  \bibinfo {author} {\bibfnamefont {R.}~\bibnamefont {Babbush}}, \bibinfo
  {author} {\bibfnamefont {D.}~\bibnamefont {Bacon}}, \bibinfo {author}
  {\bibfnamefont {J.~C.}\ \bibnamefont {Bardin}}, \bibinfo {author}
  {\bibfnamefont {G.}~\bibnamefont {Bortoli}}, \bibinfo {author} {\bibfnamefont
  {A.}~\bibnamefont {Bourassa}}, \bibinfo {author} {\bibfnamefont
  {J.}~\bibnamefont {Bovaird}}, \bibinfo {author} {\bibfnamefont
  {L.}~\bibnamefont {Brill}}, \bibinfo {author} {\bibfnamefont
  {M.}~\bibnamefont {Broughton}}, \bibinfo {author} {\bibfnamefont {B.~B.}\
  \bibnamefont {Buckley}}, \bibinfo {author} {\bibfnamefont {D.~A.}\
  \bibnamefont {Buell}}, \bibinfo {author} {\bibfnamefont {T.}~\bibnamefont
  {Burger}}, \bibinfo {author} {\bibfnamefont {B.}~\bibnamefont {Burkett}},
  \bibinfo {author} {\bibfnamefont {N.}~\bibnamefont {Bushnell}}, \bibinfo
  {author} {\bibfnamefont {Z.}~\bibnamefont {Chen}}, \bibinfo {author}
  {\bibfnamefont {B.}~\bibnamefont {Chiaro}}, \bibinfo {author} {\bibfnamefont
  {D.}~\bibnamefont {Chik}}, \bibinfo {author} {\bibfnamefont {C.}~\bibnamefont
  {Chou}}, \bibinfo {author} {\bibfnamefont {J.}~\bibnamefont {Cogan}},
  \bibinfo {author} {\bibfnamefont {R.}~\bibnamefont {Collins}}, \bibinfo
  {author} {\bibfnamefont {P.}~\bibnamefont {Conner}}, \bibinfo {author}
  {\bibfnamefont {W.}~\bibnamefont {Courtney}}, \bibinfo {author}
  {\bibfnamefont {A.~L.}\ \bibnamefont {Crook}}, \bibinfo {author}
  {\bibfnamefont {B.}~\bibnamefont {Curtin}}, \bibinfo {author} {\bibfnamefont
  {D.~M.}\ \bibnamefont {Debroy}}, \bibinfo {author} {\bibfnamefont {A.~D.~T.}\
  \bibnamefont {Barba}}, \bibinfo {author} {\bibfnamefont {S.}~\bibnamefont
  {Demura}}, \bibinfo {author} {\bibfnamefont {A.}~\bibnamefont {Dunsworth}},
  \bibinfo {author} {\bibfnamefont {D.}~\bibnamefont {Eppens}}, \bibinfo
  {author} {\bibfnamefont {C.}~\bibnamefont {Erickson}}, \bibinfo {author}
  {\bibfnamefont {L.}~\bibnamefont {Faoro}}, \bibinfo {author} {\bibfnamefont
  {E.}~\bibnamefont {Farhi}}, \bibinfo {author} {\bibfnamefont
  {R.}~\bibnamefont {Fatemi}}, \bibinfo {author} {\bibfnamefont {V.~S.}\
  \bibnamefont {Ferreira}}, \bibinfo {author} {\bibfnamefont {L.~F.}\
  \bibnamefont {Burgos}}, \bibinfo {author} {\bibfnamefont {E.}~\bibnamefont
  {Forati}}, \bibinfo {author} {\bibfnamefont {A.~G.}\ \bibnamefont {Fowler}},
  \bibinfo {author} {\bibfnamefont {B.}~\bibnamefont {Foxen}}, \bibinfo
  {author} {\bibfnamefont {W.}~\bibnamefont {Giang}}, \bibinfo {author}
  {\bibfnamefont {C.}~\bibnamefont {Gidney}}, \bibinfo {author} {\bibfnamefont
  {D.}~\bibnamefont {Gilboa}}, \bibinfo {author} {\bibfnamefont
  {M.}~\bibnamefont {Giustina}}, \bibinfo {author} {\bibfnamefont
  {R.}~\bibnamefont {Gosula}}, \bibinfo {author} {\bibfnamefont {A.~G.}\
  \bibnamefont {Dau}}, \bibinfo {author} {\bibfnamefont {J.~A.}\ \bibnamefont
  {Gross}}, \bibinfo {author} {\bibfnamefont {S.}~\bibnamefont {Habegger}},
  \bibinfo {author} {\bibfnamefont {M.~C.}\ \bibnamefont {Hamilton}}, \bibinfo
  {author} {\bibfnamefont {M.}~\bibnamefont {Hansen}}, \bibinfo {author}
  {\bibfnamefont {M.~P.}\ \bibnamefont {Harrigan}}, \bibinfo {author}
  {\bibfnamefont {S.~D.}\ \bibnamefont {Harrington}}, \bibinfo {author}
  {\bibfnamefont {P.}~\bibnamefont {Heu}}, \bibinfo {author} {\bibfnamefont
  {J.}~\bibnamefont {Hilton}}, \bibinfo {author} {\bibfnamefont {M.~R.}\
  \bibnamefont {Hoffmann}}, \bibinfo {author} {\bibfnamefont {T.}~\bibnamefont
  {Huang}}, \bibinfo {author} {\bibfnamefont {A.}~\bibnamefont {Huff}},
  \bibinfo {author} {\bibfnamefont {W.~J.}\ \bibnamefont {Huggins}}, \bibinfo
  {author} {\bibfnamefont {L.~B.}\ \bibnamefont {Ioffe}}, \bibinfo {author}
  {\bibfnamefont {S.~V.}\ \bibnamefont {Isakov}}, \bibinfo {author}
  {\bibfnamefont {J.}~\bibnamefont {Iveland}}, \bibinfo {author} {\bibfnamefont
  {E.}~\bibnamefont {Jeffrey}}, \bibinfo {author} {\bibfnamefont
  {Z.}~\bibnamefont {Jiang}}, \bibinfo {author} {\bibfnamefont
  {C.}~\bibnamefont {Jones}}, \bibinfo {author} {\bibfnamefont
  {P.}~\bibnamefont {Juhas}}, \bibinfo {author} {\bibfnamefont
  {D.}~\bibnamefont {Kafri}}, \bibinfo {author} {\bibfnamefont
  {T.}~\bibnamefont {Khattar}}, \bibinfo {author} {\bibfnamefont
  {M.}~\bibnamefont {Khezri}}, \bibinfo {author} {\bibfnamefont
  {M.}~\bibnamefont {Kieferová}}, \bibinfo {author} {\bibfnamefont
  {S.}~\bibnamefont {Kim}}, \bibinfo {author} {\bibfnamefont {A.}~\bibnamefont
  {Kitaev}}, \bibinfo {author} {\bibfnamefont {P.~V.}\ \bibnamefont {Klimov}},
  \bibinfo {author} {\bibfnamefont {A.~R.}\ \bibnamefont {Klots}}, \bibinfo
  {author} {\bibfnamefont {A.~N.}\ \bibnamefont {Korotkov}}, \bibinfo {author}
  {\bibfnamefont {F.}~\bibnamefont {Kostritsa}}, \bibinfo {author}
  {\bibfnamefont {J.~M.}\ \bibnamefont {Kreikebaum}}, \bibinfo {author}
  {\bibfnamefont {D.}~\bibnamefont {Landhuis}}, \bibinfo {author}
  {\bibfnamefont {P.}~\bibnamefont {Laptev}}, \bibinfo {author} {\bibfnamefont
  {K.-M.}\ \bibnamefont {Lau}}, \bibinfo {author} {\bibfnamefont
  {L.}~\bibnamefont {Laws}}, \bibinfo {author} {\bibfnamefont {J.}~\bibnamefont
  {Lee}}, \bibinfo {author} {\bibfnamefont {K.}~\bibnamefont {Lee}}, \bibinfo
  {author} {\bibfnamefont {B.~J.}\ \bibnamefont {Lester}}, \bibinfo {author}
  {\bibfnamefont {A.}~\bibnamefont {Lill}}, \bibinfo {author} {\bibfnamefont
  {W.}~\bibnamefont {Liu}}, \bibinfo {author} {\bibfnamefont {A.}~\bibnamefont
  {Locharla}}, \bibinfo {author} {\bibfnamefont {E.}~\bibnamefont {Lucero}},
  \bibinfo {author} {\bibfnamefont {F.~D.}\ \bibnamefont {Malone}}, \bibinfo
  {author} {\bibfnamefont {O.}~\bibnamefont {Martin}}, \bibinfo {author}
  {\bibfnamefont {J.~R.}\ \bibnamefont {McClean}}, \bibinfo {author}
  {\bibfnamefont {T.}~\bibnamefont {McCourt}}, \bibinfo {author} {\bibfnamefont
  {M.}~\bibnamefont {McEwen}}, \bibinfo {author} {\bibfnamefont {K.~C.}\
  \bibnamefont {Miao}}, \bibinfo {author} {\bibfnamefont {A.}~\bibnamefont
  {Mieszala}}, \bibinfo {author} {\bibfnamefont {M.}~\bibnamefont {Mohseni}},
  \bibinfo {author} {\bibfnamefont {S.}~\bibnamefont {Montazeri}}, \bibinfo
  {author} {\bibfnamefont {E.}~\bibnamefont {Mount}}, \bibinfo {author}
  {\bibfnamefont {R.}~\bibnamefont {Movassagh}}, \bibinfo {author}
  {\bibfnamefont {W.}~\bibnamefont {Mruczkiewicz}}, \bibinfo {author}
  {\bibfnamefont {O.}~\bibnamefont {Naaman}}, \bibinfo {author} {\bibfnamefont
  {M.}~\bibnamefont {Neeley}}, \bibinfo {author} {\bibfnamefont
  {C.}~\bibnamefont {Neill}}, \bibinfo {author} {\bibfnamefont
  {A.}~\bibnamefont {Nersisyan}}, \bibinfo {author} {\bibfnamefont
  {M.}~\bibnamefont {Newman}}, \bibinfo {author} {\bibfnamefont {J.~H.}\
  \bibnamefont {Ng}}, \bibinfo {author} {\bibfnamefont {A.}~\bibnamefont
  {Nguyen}}, \bibinfo {author} {\bibfnamefont {M.}~\bibnamefont {Nguyen}},
  \bibinfo {author} {\bibfnamefont {M.~Y.}\ \bibnamefont {Niu}}, \bibinfo
  {author} {\bibfnamefont {T.~E.}\ \bibnamefont {O'Brien}}, \bibinfo {author}
  {\bibfnamefont {S.}~\bibnamefont {Omonije}}, \bibinfo {author} {\bibfnamefont
  {A.}~\bibnamefont {Petukhov}}, \bibinfo {author} {\bibfnamefont
  {R.}~\bibnamefont {Potter}}, \bibinfo {author} {\bibfnamefont {L.~P.}\
  \bibnamefont {Pryadko}}, \bibinfo {author} {\bibfnamefont {C.}~\bibnamefont
  {Quintana}}, \bibinfo {author} {\bibfnamefont {C.}~\bibnamefont {Rocque}},
  \bibinfo {author} {\bibfnamefont {N.~C.}\ \bibnamefont {Rubin}}, \bibinfo
  {author} {\bibfnamefont {N.}~\bibnamefont {Saei}}, \bibinfo {author}
  {\bibfnamefont {D.}~\bibnamefont {Sank}}, \bibinfo {author} {\bibfnamefont
  {K.}~\bibnamefont {Sankaragomathi}}, \bibinfo {author} {\bibfnamefont
  {K.~J.}\ \bibnamefont {Satzinger}}, \bibinfo {author} {\bibfnamefont {H.~F.}\
  \bibnamefont {Schurkus}}, \bibinfo {author} {\bibfnamefont {C.}~\bibnamefont
  {Schuster}}, \bibinfo {author} {\bibfnamefont {M.~J.}\ \bibnamefont
  {Shearn}}, \bibinfo {author} {\bibfnamefont {A.}~\bibnamefont {Shorter}},
  \bibinfo {author} {\bibfnamefont {N.}~\bibnamefont {Shutty}}, \bibinfo
  {author} {\bibfnamefont {V.}~\bibnamefont {Shvarts}}, \bibinfo {author}
  {\bibfnamefont {J.}~\bibnamefont {Skruzny}}, \bibinfo {author} {\bibfnamefont
  {W.~C.}\ \bibnamefont {Smith}}, \bibinfo {author} {\bibfnamefont
  {R.}~\bibnamefont {Somma}}, \bibinfo {author} {\bibfnamefont
  {G.}~\bibnamefont {Sterling}}, \bibinfo {author} {\bibfnamefont
  {D.}~\bibnamefont {Strain}}, \bibinfo {author} {\bibfnamefont
  {M.}~\bibnamefont {Szalay}}, \bibinfo {author} {\bibfnamefont
  {A.}~\bibnamefont {Torres}}, \bibinfo {author} {\bibfnamefont
  {G.}~\bibnamefont {Vidal}}, \bibinfo {author} {\bibfnamefont
  {B.}~\bibnamefont {Villalonga}}, \bibinfo {author} {\bibfnamefont {C.~V.}\
  \bibnamefont {Heidweiller}}, \bibinfo {author} {\bibfnamefont
  {T.}~\bibnamefont {White}}, \bibinfo {author} {\bibfnamefont {B.~W.~K.}\
  \bibnamefont {Woo}}, \bibinfo {author} {\bibfnamefont {C.}~\bibnamefont
  {Xing}}, \bibinfo {author} {\bibfnamefont {Z.~J.}\ \bibnamefont {Yao}},
  \bibinfo {author} {\bibfnamefont {P.}~\bibnamefont {Yeh}}, \bibinfo {author}
  {\bibfnamefont {J.}~\bibnamefont {Yoo}}, \bibinfo {author} {\bibfnamefont
  {G.}~\bibnamefont {Young}}, \bibinfo {author} {\bibfnamefont
  {A.}~\bibnamefont {Zalcman}}, \bibinfo {author} {\bibfnamefont
  {Y.}~\bibnamefont {Zhang}}, \bibinfo {author} {\bibfnamefont
  {N.}~\bibnamefont {Zhu}}, \bibinfo {author} {\bibfnamefont {N.}~\bibnamefont
  {Zobrist}}, \bibinfo {author} {\bibfnamefont {H.}~\bibnamefont {Neven}},
  \bibinfo {author} {\bibfnamefont {S.}~\bibnamefont {Boixo}}, \bibinfo
  {author} {\bibfnamefont {A.}~\bibnamefont {Megrant}}, \bibinfo {author}
  {\bibfnamefont {J.}~\bibnamefont {Kelly}}, \bibinfo {author} {\bibfnamefont
  {Y.}~\bibnamefont {Chen}}, \bibinfo {author} {\bibfnamefont {V.}~\bibnamefont
  {Smelyanskiy}}, \bibinfo {author} {\bibfnamefont {E.-A.}\ \bibnamefont
  {Kim}}, \bibinfo {author} {\bibfnamefont {I.}~\bibnamefont {Aleiner}},\ and\
  \bibinfo {author} {\bibfnamefont {P.}~\bibnamefont {Roushan}},\ }\href
  {https://doi.org/10.48550/arXiv.2210.10255} {\bibinfo {title} {Observation of
  non-{Abelian} exchange statistics on a superconducting processor}} (\bibinfo
  {year} {2022})\BibitemShut {NoStop}%
\bibitem [{\citenamefont {Xu}\ \emph {et~al.}(2022)\citenamefont {Xu},
  \citenamefont {Sun}, \citenamefont {Wang}, \citenamefont {Xiang},
  \citenamefont {Bao}, \citenamefont {Zhu}, \citenamefont {Shen}, \citenamefont
  {Song}, \citenamefont {Zhang}, \citenamefont {Ren}, \citenamefont {Zhang},
  \citenamefont {Dong}, \citenamefont {Deng}, \citenamefont {Chen},
  \citenamefont {Wu}, \citenamefont {Tan}, \citenamefont {Gao}, \citenamefont
  {Jin}, \citenamefont {Zhu}, \citenamefont {Zhang}, \citenamefont {Wang},
  \citenamefont {Zou}, \citenamefont {Zhong}, \citenamefont {Zhang},
  \citenamefont {Li}, \citenamefont {Jiang}, \citenamefont {Yu}, \citenamefont
  {Yao}, \citenamefont {Wang}, \citenamefont {Li}, \citenamefont {Guo},
  \citenamefont {Song}, \citenamefont {Wang},\ and\ \citenamefont
  {Deng}}]{xu_digital_2022}%
  \BibitemOpen
  \bibfield  {author} {\bibinfo {author} {\bibfnamefont {S.}~\bibnamefont
  {Xu}}, \bibinfo {author} {\bibfnamefont {Z.-Z.}\ \bibnamefont {Sun}},
  \bibinfo {author} {\bibfnamefont {K.}~\bibnamefont {Wang}}, \bibinfo {author}
  {\bibfnamefont {L.}~\bibnamefont {Xiang}}, \bibinfo {author} {\bibfnamefont
  {Z.}~\bibnamefont {Bao}}, \bibinfo {author} {\bibfnamefont {Z.}~\bibnamefont
  {Zhu}}, \bibinfo {author} {\bibfnamefont {F.}~\bibnamefont {Shen}}, \bibinfo
  {author} {\bibfnamefont {Z.}~\bibnamefont {Song}}, \bibinfo {author}
  {\bibfnamefont {P.}~\bibnamefont {Zhang}}, \bibinfo {author} {\bibfnamefont
  {W.}~\bibnamefont {Ren}}, \bibinfo {author} {\bibfnamefont {X.}~\bibnamefont
  {Zhang}}, \bibinfo {author} {\bibfnamefont {H.}~\bibnamefont {Dong}},
  \bibinfo {author} {\bibfnamefont {J.}~\bibnamefont {Deng}}, \bibinfo {author}
  {\bibfnamefont {J.}~\bibnamefont {Chen}}, \bibinfo {author} {\bibfnamefont
  {Y.}~\bibnamefont {Wu}}, \bibinfo {author} {\bibfnamefont {Z.}~\bibnamefont
  {Tan}}, \bibinfo {author} {\bibfnamefont {Y.}~\bibnamefont {Gao}}, \bibinfo
  {author} {\bibfnamefont {F.}~\bibnamefont {Jin}}, \bibinfo {author}
  {\bibfnamefont {X.}~\bibnamefont {Zhu}}, \bibinfo {author} {\bibfnamefont
  {C.}~\bibnamefont {Zhang}}, \bibinfo {author} {\bibfnamefont
  {N.}~\bibnamefont {Wang}}, \bibinfo {author} {\bibfnamefont {Y.}~\bibnamefont
  {Zou}}, \bibinfo {author} {\bibfnamefont {J.}~\bibnamefont {Zhong}}, \bibinfo
  {author} {\bibfnamefont {A.}~\bibnamefont {Zhang}}, \bibinfo {author}
  {\bibfnamefont {W.}~\bibnamefont {Li}}, \bibinfo {author} {\bibfnamefont
  {W.}~\bibnamefont {Jiang}}, \bibinfo {author} {\bibfnamefont {L.-W.}\
  \bibnamefont {Yu}}, \bibinfo {author} {\bibfnamefont {Y.}~\bibnamefont
  {Yao}}, \bibinfo {author} {\bibfnamefont {Z.}~\bibnamefont {Wang}}, \bibinfo
  {author} {\bibfnamefont {H.}~\bibnamefont {Li}}, \bibinfo {author}
  {\bibfnamefont {Q.}~\bibnamefont {Guo}}, \bibinfo {author} {\bibfnamefont
  {C.}~\bibnamefont {Song}}, \bibinfo {author} {\bibfnamefont {H.}~\bibnamefont
  {Wang}},\ and\ \bibinfo {author} {\bibfnamefont {D.-L.}\ \bibnamefont
  {Deng}},\ }\href {https://doi.org/10.48550/arXiv.2211.09802} {\bibinfo
  {title} {Digital simulation of non-{Abelian} anyons with 68 programmable
  superconducting qubits}} (\bibinfo {year} {2022})\BibitemShut {NoStop}%
\bibitem [{\citenamefont {Krinner}\ \emph {et~al.}(2022)\citenamefont
  {Krinner}, \citenamefont {Lacroix}, \citenamefont {Remm}, \citenamefont
  {Di~Paolo}, \citenamefont {Genois}, \citenamefont {Leroux}, \citenamefont
  {Hellings}, \citenamefont {Lazar}, \citenamefont {Swiadek}, \citenamefont
  {Herrmann}, \citenamefont {Norris}, \citenamefont {Andersen}, \citenamefont
  {Müller}, \citenamefont {Blais}, \citenamefont {Eichler},\ and\
  \citenamefont {Wallraff}}]{krinner_realizing_2022}%
  \BibitemOpen
  \bibfield  {author} {\bibinfo {author} {\bibfnamefont {S.}~\bibnamefont
  {Krinner}}, \bibinfo {author} {\bibfnamefont {N.}~\bibnamefont {Lacroix}},
  \bibinfo {author} {\bibfnamefont {A.}~\bibnamefont {Remm}}, \bibinfo {author}
  {\bibfnamefont {A.}~\bibnamefont {Di~Paolo}}, \bibinfo {author}
  {\bibfnamefont {E.}~\bibnamefont {Genois}}, \bibinfo {author} {\bibfnamefont
  {C.}~\bibnamefont {Leroux}}, \bibinfo {author} {\bibfnamefont
  {C.}~\bibnamefont {Hellings}}, \bibinfo {author} {\bibfnamefont
  {S.}~\bibnamefont {Lazar}}, \bibinfo {author} {\bibfnamefont
  {F.}~\bibnamefont {Swiadek}}, \bibinfo {author} {\bibfnamefont
  {J.}~\bibnamefont {Herrmann}}, \bibinfo {author} {\bibfnamefont {G.~J.}\
  \bibnamefont {Norris}}, \bibinfo {author} {\bibfnamefont {C.~K.}\
  \bibnamefont {Andersen}}, \bibinfo {author} {\bibfnamefont {M.}~\bibnamefont
  {Müller}}, \bibinfo {author} {\bibfnamefont {A.}~\bibnamefont {Blais}},
  \bibinfo {author} {\bibfnamefont {C.}~\bibnamefont {Eichler}},\ and\ \bibinfo
  {author} {\bibfnamefont {A.}~\bibnamefont {Wallraff}},\ }\bibfield  {title}
  {\bibinfo {title} {Realizing {Repeated} {Quantum} {Error} {Correction} in a
  {Distance}-{Three} {Surface} {Code}},\ }\href
  {https://doi.org/10.1038/s41586-022-04566-8} {\bibfield  {journal} {\bibinfo
  {journal} {Nature}\ }\textbf {\bibinfo {volume} {605}},\ \bibinfo {pages}
  {669} (\bibinfo {year} {2022})},\ \bibinfo {note} {arXiv:2112.03708
  [cond-mat, physics:quant-ph]}\BibitemShut {NoStop}%
\bibitem [{\citenamefont {Gottesman}(1997)}]{gottesman_stabilizer_1997}%
  \BibitemOpen
  \bibfield  {author} {\bibinfo {author} {\bibfnamefont {D.}~\bibnamefont
  {Gottesman}},\ }\href {https://doi.org/10.48550/arXiv.quant-ph/9705052}
  {\bibinfo {title} {Stabilizer {Codes} and {Quantum} {Error} {Correction}}}
  (\bibinfo {year} {1997})\BibitemShut {NoStop}%
\bibitem [{\citenamefont {Raussendorf}\ \emph {et~al.}(2005)\citenamefont
  {Raussendorf}, \citenamefont {Bravyi},\ and\ \citenamefont
  {Harrington}}]{raussendorf_long-range_2005}%
  \BibitemOpen
  \bibfield  {author} {\bibinfo {author} {\bibfnamefont {R.}~\bibnamefont
  {Raussendorf}}, \bibinfo {author} {\bibfnamefont {S.}~\bibnamefont
  {Bravyi}},\ and\ \bibinfo {author} {\bibfnamefont {J.}~\bibnamefont
  {Harrington}},\ }\bibfield  {title} {\bibinfo {title} {Long-range quantum
  entanglement in noisy cluster states},\ }\href
  {https://doi.org/10.1103/PhysRevA.71.062313} {\bibfield  {journal} {\bibinfo
  {journal} {Physical Review A}\ }\textbf {\bibinfo {volume} {71}},\ \bibinfo
  {pages} {062313} (\bibinfo {year} {2005})}\BibitemShut {NoStop}%
\bibitem [{\citenamefont {Pino}\ \emph {et~al.}(2021)\citenamefont {Pino},
  \citenamefont {Dreiling}, \citenamefont {Figgatt}, \citenamefont {Gaebler},
  \citenamefont {Moses}, \citenamefont {Allman}, \citenamefont {Baldwin},
  \citenamefont {Foss-Feig}, \citenamefont {Hayes}, \citenamefont {Mayer},
  \citenamefont {Ryan-Anderson},\ and\ \citenamefont
  {Neyenhuis}}]{pino_demonstration_2021}%
  \BibitemOpen
  \bibfield  {author} {\bibinfo {author} {\bibfnamefont {J.~M.}\ \bibnamefont
  {Pino}}, \bibinfo {author} {\bibfnamefont {J.~M.}\ \bibnamefont {Dreiling}},
  \bibinfo {author} {\bibfnamefont {C.}~\bibnamefont {Figgatt}}, \bibinfo
  {author} {\bibfnamefont {J.~P.}\ \bibnamefont {Gaebler}}, \bibinfo {author}
  {\bibfnamefont {S.~A.}\ \bibnamefont {Moses}}, \bibinfo {author}
  {\bibfnamefont {M.~S.}\ \bibnamefont {Allman}}, \bibinfo {author}
  {\bibfnamefont {C.~H.}\ \bibnamefont {Baldwin}}, \bibinfo {author}
  {\bibfnamefont {M.}~\bibnamefont {Foss-Feig}}, \bibinfo {author}
  {\bibfnamefont {D.}~\bibnamefont {Hayes}}, \bibinfo {author} {\bibfnamefont
  {K.}~\bibnamefont {Mayer}}, \bibinfo {author} {\bibfnamefont
  {C.}~\bibnamefont {Ryan-Anderson}},\ and\ \bibinfo {author} {\bibfnamefont
  {B.}~\bibnamefont {Neyenhuis}},\ }\bibfield  {title} {\bibinfo {title}
  {Demonstration of the trapped-ion quantum-{CCD} computer architecture},\
  }\href {https://doi.org/10.1038/s41586-021-03318-4} {\bibfield  {journal}
  {\bibinfo  {journal} {Nature}\ }\textbf {\bibinfo {volume} {592}},\ \bibinfo
  {pages} {209} (\bibinfo {year} {2021})}\BibitemShut {NoStop}%
\bibitem [{\citenamefont {Bombin}(2010)}]{Bombin10Defect}%
  \BibitemOpen
  \bibfield  {author} {\bibinfo {author} {\bibfnamefont {H.}~\bibnamefont
  {Bombin}},\ }\bibfield  {title} {\bibinfo {title} {Topological order with a
  twist: Ising anyons from an abelian model},\ }\href
  {https://doi.org/10.1103/PhysRevLett.105.030403} {\bibfield  {journal}
  {\bibinfo  {journal} {Phys. Rev. Lett.}\ }\textbf {\bibinfo {volume} {105}},\
  \bibinfo {pages} {030403} (\bibinfo {year} {2010})}\BibitemShut {NoStop}%
\bibitem [{\citenamefont {Kitaev}\ and\ \citenamefont
  {Kong}(2012)}]{KitaevKong12}%
  \BibitemOpen
  \bibfield  {author} {\bibinfo {author} {\bibfnamefont {A.}~\bibnamefont
  {Kitaev}}\ and\ \bibinfo {author} {\bibfnamefont {L.}~\bibnamefont {Kong}},\
  }\bibfield  {title} {\bibinfo {title} {Models for gapped boundaries and
  domain walls},\ }\href {https://doi.org/10.1007/s00220-012-1500-5} {\bibfield
   {journal} {\bibinfo  {journal} {Communications in Mathematical Physics}\
  }\textbf {\bibinfo {volume} {313}},\ \bibinfo {pages} {351} (\bibinfo {year}
  {2012})}\BibitemShut {NoStop}%
\bibitem [{\citenamefont {Kitaev}(2003)}]{kitaev_fault-tolerant_2003}%
  \BibitemOpen
  \bibfield  {author} {\bibinfo {author} {\bibfnamefont {A.~Y.}\ \bibnamefont
  {Kitaev}},\ }\bibfield  {title} {\bibinfo {title} {Fault-tolerant quantum
  computation by anyons},\ }\href
  {https://doi.org/10.1016/S0003-4916(02)00018-0} {\bibfield  {journal}
  {\bibinfo  {journal} {Annals of Physics}\ }\textbf {\bibinfo {volume}
  {303}},\ \bibinfo {pages} {2} (\bibinfo {year} {2003})}\BibitemShut {NoStop}%
\bibitem [{\citenamefont {Wen}(2003)}]{Wen_plaquette}%
  \BibitemOpen
  \bibfield  {author} {\bibinfo {author} {\bibfnamefont {X.-G.}\ \bibnamefont
  {Wen}},\ }\bibfield  {title} {\bibinfo {title} {Quantum orders in an exact
  soluble model},\ }\href {https://doi.org/10.1103/PhysRevLett.90.016803}
  {\bibfield  {journal} {\bibinfo  {journal} {Phys. Rev. Lett.}\ }\textbf
  {\bibinfo {volume} {90}},\ \bibinfo {pages} {016803} (\bibinfo {year}
  {2003})}\BibitemShut {NoStop}%
\bibitem [{\citenamefont {Read}\ and\ \citenamefont {Sachdev}(1991)}]{Read91}%
  \BibitemOpen
  \bibfield  {author} {\bibinfo {author} {\bibfnamefont {N.}~\bibnamefont
  {Read}}\ and\ \bibinfo {author} {\bibfnamefont {S.}~\bibnamefont {Sachdev}},\
  }\bibfield  {title} {\bibinfo {title} {Large-n expansion for frustrated
  quantum antiferromagnets},\ }\href
  {https://doi.org/10.1103/PhysRevLett.66.1773} {\bibfield  {journal} {\bibinfo
   {journal} {Phys. Rev. Lett.}\ }\textbf {\bibinfo {volume} {66}},\ \bibinfo
  {pages} {1773} (\bibinfo {year} {1991})}\BibitemShut {NoStop}%
\bibitem [{\citenamefont {Wen}(1991)}]{Wen91}%
  \BibitemOpen
  \bibfield  {author} {\bibinfo {author} {\bibfnamefont {X.~G.}\ \bibnamefont
  {Wen}},\ }\bibfield  {title} {\bibinfo {title} {Mean-field theory of
  spin-liquid states with finite energy gap and topological orders},\ }\href
  {https://doi.org/10.1103/PhysRevB.44.2664} {\bibfield  {journal} {\bibinfo
  {journal} {Phys. Rev. B}\ }\textbf {\bibinfo {volume} {44}},\ \bibinfo
  {pages} {2664} (\bibinfo {year} {1991})}\BibitemShut {NoStop}%
\bibitem [{\citenamefont {Cramer}\ \emph {et~al.}(2010)\citenamefont {Cramer},
  \citenamefont {Plenio}, \citenamefont {Flammia}, \citenamefont {Somma},
  \citenamefont {Gross}, \citenamefont {Bartlett}, \citenamefont
  {Landon-Cardinal}, \citenamefont {Poulin},\ and\ \citenamefont
  {Liu}}]{cramer_2010}%
  \BibitemOpen
  \bibfield  {author} {\bibinfo {author} {\bibfnamefont {M.}~\bibnamefont
  {Cramer}}, \bibinfo {author} {\bibfnamefont {M.~B.}\ \bibnamefont {Plenio}},
  \bibinfo {author} {\bibfnamefont {S.~T.}\ \bibnamefont {Flammia}}, \bibinfo
  {author} {\bibfnamefont {R.}~\bibnamefont {Somma}}, \bibinfo {author}
  {\bibfnamefont {D.}~\bibnamefont {Gross}}, \bibinfo {author} {\bibfnamefont
  {S.~D.}\ \bibnamefont {Bartlett}}, \bibinfo {author} {\bibfnamefont
  {O.}~\bibnamefont {Landon-Cardinal}}, \bibinfo {author} {\bibfnamefont
  {D.}~\bibnamefont {Poulin}},\ and\ \bibinfo {author} {\bibfnamefont {Y.-K.}\
  \bibnamefont {Liu}},\ }\bibfield  {title} {\bibinfo {title} {Efficient
  quantum state tomography},\ }\href {https://doi.org/10.1038/ncomms1147}
  {\bibfield  {journal} {\bibinfo  {journal} {Nature Communications}\ }\textbf
  {\bibinfo {volume} {1}},\ \bibinfo {pages} {149} (\bibinfo {year}
  {2010})}\BibitemShut {NoStop}%
\bibitem [{\citenamefont {Kitaev}\ and\ \citenamefont
  {Preskill}(2006)}]{kitaev_topological_2006}%
  \BibitemOpen
  \bibfield  {author} {\bibinfo {author} {\bibfnamefont {A.}~\bibnamefont
  {Kitaev}}\ and\ \bibinfo {author} {\bibfnamefont {J.}~\bibnamefont
  {Preskill}},\ }\bibfield  {title} {\bibinfo {title} {Topological entanglement
  entropy},\ }\href {https://doi.org/10.1103/PhysRevLett.96.110404} {\bibfield
  {journal} {\bibinfo  {journal} {Physical Review Letters}\ }\textbf {\bibinfo
  {volume} {96}},\ \bibinfo {pages} {110404} (\bibinfo {year}
  {2006})}\BibitemShut {NoStop}%
\bibitem [{\citenamefont {Levin}\ and\ \citenamefont
  {Wen}(2006)}]{levin_detecting_2006}%
  \BibitemOpen
  \bibfield  {author} {\bibinfo {author} {\bibfnamefont {M.}~\bibnamefont
  {Levin}}\ and\ \bibinfo {author} {\bibfnamefont {X.-G.}\ \bibnamefont
  {Wen}},\ }\bibfield  {title} {\bibinfo {title} {Detecting topological order
  in a ground state wave function},\ }\href
  {https://doi.org/10.1103/PhysRevLett.96.110405} {\bibfield  {journal}
  {\bibinfo  {journal} {Physical Review Letters}\ }\textbf {\bibinfo {volume}
  {96}},\ \bibinfo {pages} {110405} (\bibinfo {year} {2006})}\BibitemShut
  {NoStop}%
\bibitem [{\citenamefont {Flammia}\ \emph {et~al.}(2009)\citenamefont
  {Flammia}, \citenamefont {Hamma}, \citenamefont {Hughes},\ and\ \citenamefont
  {Wen}}]{flammia_topological_2009}%
  \BibitemOpen
  \bibfield  {author} {\bibinfo {author} {\bibfnamefont {S.~T.}\ \bibnamefont
  {Flammia}}, \bibinfo {author} {\bibfnamefont {A.}~\bibnamefont {Hamma}},
  \bibinfo {author} {\bibfnamefont {T.~L.}\ \bibnamefont {Hughes}},\ and\
  \bibinfo {author} {\bibfnamefont {X.-G.}\ \bibnamefont {Wen}},\ }\bibfield
  {title} {\bibinfo {title} {Topological {Entanglement} {Renyi} {Entropy} and
  {Reduced} {Density} {Matrix} {Structure}},\ }\href
  {https://doi.org/10.1103/PhysRevLett.103.261601} {\bibfield  {journal}
  {\bibinfo  {journal} {Physical Review Letters}\ }\textbf {\bibinfo {volume}
  {103}},\ \bibinfo {pages} {261601} (\bibinfo {year} {2009})}\BibitemShut
  {NoStop}%
\bibitem [{\citenamefont {van Enk}\ and\ \citenamefont
  {Beenakker}(2012)}]{van2011power}%
  \BibitemOpen
  \bibfield  {author} {\bibinfo {author} {\bibfnamefont {S.~J.}\ \bibnamefont
  {van Enk}}\ and\ \bibinfo {author} {\bibfnamefont {C.~W.~J.}\ \bibnamefont
  {Beenakker}},\ }\bibfield  {title} {\bibinfo {title} {Measuring
  $\mathrm{Tr}{\ensuremath{\rho}}^{n}$ on single copies of $\ensuremath{\rho}$
  using random measurements},\ }\href
  {https://doi.org/10.1103/PhysRevLett.108.110503} {\bibfield  {journal}
  {\bibinfo  {journal} {Phys. Rev. Lett.}\ }\textbf {\bibinfo {volume} {108}},\
  \bibinfo {pages} {110503} (\bibinfo {year} {2012})}\BibitemShut {NoStop}%
\bibitem [{\citenamefont {Elben}\ \emph {et~al.}(2018)\citenamefont {Elben},
  \citenamefont {Vermersch}, \citenamefont {Dalmonte}, \citenamefont {Cirac},\
  and\ \citenamefont {Zoller}}]{elben_renyi_2018}%
  \BibitemOpen
  \bibfield  {author} {\bibinfo {author} {\bibfnamefont {A.}~\bibnamefont
  {Elben}}, \bibinfo {author} {\bibfnamefont {B.}~\bibnamefont {Vermersch}},
  \bibinfo {author} {\bibfnamefont {M.}~\bibnamefont {Dalmonte}}, \bibinfo
  {author} {\bibfnamefont {J.}~\bibnamefont {Cirac}},\ and\ \bibinfo {author}
  {\bibfnamefont {P.}~\bibnamefont {Zoller}},\ }\bibfield  {title} {\bibinfo
  {title} {Renyi {Entropies} from {Random} {Quenches} in {Atomic} {Hubbard} and
  {Spin} {Models}},\ }\href {https://doi.org/10.1103/PhysRevLett.120.050406}
  {\bibfield  {journal} {\bibinfo  {journal} {Physical Review Letters}\
  }\textbf {\bibinfo {volume} {120}},\ \bibinfo {pages} {050406} (\bibinfo
  {year} {2018})}\BibitemShut {NoStop}%
\bibitem [{\citenamefont {Vermersch}\ \emph {et~al.}(2018)\citenamefont
  {Vermersch}, \citenamefont {Elben}, \citenamefont {Dalmonte}, \citenamefont
  {Cirac},\ and\ \citenamefont {Zoller}}]{vermersch_unitary_2018}%
  \BibitemOpen
  \bibfield  {author} {\bibinfo {author} {\bibfnamefont {B.}~\bibnamefont
  {Vermersch}}, \bibinfo {author} {\bibfnamefont {A.}~\bibnamefont {Elben}},
  \bibinfo {author} {\bibfnamefont {M.}~\bibnamefont {Dalmonte}}, \bibinfo
  {author} {\bibfnamefont {J.~I.}\ \bibnamefont {Cirac}},\ and\ \bibinfo
  {author} {\bibfnamefont {P.}~\bibnamefont {Zoller}},\ }\bibfield  {title}
  {\bibinfo {title} {Unitary n-designs via random quenches in atomic {Hubbard}
  and spin models: {Application} to the measurement of {Renyi} entropies},\
  }\href {https://doi.org/10.1103/PhysRevA.97.023604} {\bibfield  {journal}
  {\bibinfo  {journal} {Physical Review A}\ }\textbf {\bibinfo {volume} {97}},\
  \bibinfo {pages} {023604} (\bibinfo {year} {2018})}\BibitemShut {NoStop}%
\bibitem [{\citenamefont {Brydges}\ \emph {et~al.}(2019)\citenamefont
  {Brydges}, \citenamefont {Elben}, \citenamefont {Jurcevic}, \citenamefont
  {Vermersch}, \citenamefont {Maier}, \citenamefont {Lanyon}, \citenamefont
  {Zoller}, \citenamefont {Blatt},\ and\ \citenamefont
  {Roos}}]{brydges_probing_2019}%
  \BibitemOpen
  \bibfield  {author} {\bibinfo {author} {\bibfnamefont {T.}~\bibnamefont
  {Brydges}}, \bibinfo {author} {\bibfnamefont {A.}~\bibnamefont {Elben}},
  \bibinfo {author} {\bibfnamefont {P.}~\bibnamefont {Jurcevic}}, \bibinfo
  {author} {\bibfnamefont {B.}~\bibnamefont {Vermersch}}, \bibinfo {author}
  {\bibfnamefont {C.}~\bibnamefont {Maier}}, \bibinfo {author} {\bibfnamefont
  {B.~P.}\ \bibnamefont {Lanyon}}, \bibinfo {author} {\bibfnamefont
  {P.}~\bibnamefont {Zoller}}, \bibinfo {author} {\bibfnamefont
  {R.}~\bibnamefont {Blatt}},\ and\ \bibinfo {author} {\bibfnamefont {C.~F.}\
  \bibnamefont {Roos}},\ }\bibfield  {title} {\bibinfo {title} {Probing
  entanglement entropy via randomized measurements},\ }\href
  {https://doi.org/10.1126/science.aau4963} {\bibfield  {journal} {\bibinfo
  {journal} {Science}\ }\textbf {\bibinfo {volume} {364}},\ \bibinfo {pages}
  {260} (\bibinfo {year} {2019})}\BibitemShut {NoStop}%
\bibitem [{\citenamefont {You}\ \emph {et~al.}(2013)\citenamefont {You},
  \citenamefont {Jian},\ and\ \citenamefont {Wen}}]{you_synthetic_2013}%
  \BibitemOpen
  \bibfield  {author} {\bibinfo {author} {\bibfnamefont {Y.-Z.}\ \bibnamefont
  {You}}, \bibinfo {author} {\bibfnamefont {C.-M.}\ \bibnamefont {Jian}},\ and\
  \bibinfo {author} {\bibfnamefont {X.-G.}\ \bibnamefont {Wen}},\ }\bibfield
  {title} {\bibinfo {title} {Synthetic non-{Abelian} statistics by {Abelian}
  anyon condensation},\ }\href {https://doi.org/10.1103/PhysRevB.87.045106}
  {\bibfield  {journal} {\bibinfo  {journal} {Physical Review B}\ }\textbf
  {\bibinfo {volume} {87}},\ \bibinfo {pages} {045106} (\bibinfo {year}
  {2013})}\BibitemShut {NoStop}%
\bibitem [{h11(2022{\natexlab{a}})}]{h11}%
  \BibitemOpen
  \href@noop {} {\bibinfo {title} {{Quantinuum H1-1}}},\ \bibinfo
  {howpublished} {\url{https://www.quantinuum.com/},} (\bibinfo {year} {Nov 14
  - Dec 16, 2022}{\natexlab{a}})\BibitemShut {NoStop}%
\bibitem [{\citenamefont {Tantivasadakarn}\ \emph
  {et~al.}(2022{\natexlab{b}})\citenamefont {Tantivasadakarn}, \citenamefont
  {Verresen},\ and\ \citenamefont
  {Vishwanath}}]{tantivasadakarn_shortest_2022}%
  \BibitemOpen
  \bibfield  {author} {\bibinfo {author} {\bibfnamefont {N.}~\bibnamefont
  {Tantivasadakarn}}, \bibinfo {author} {\bibfnamefont {R.}~\bibnamefont
  {Verresen}},\ and\ \bibinfo {author} {\bibfnamefont {A.}~\bibnamefont
  {Vishwanath}},\ }\href {https://doi.org/10.48550/arXiv.2209.03964} {\bibinfo
  {title} {The {Shortest} {Route} to {Non}-{Abelian} {Topological} {Order} on a
  {Quantum} {Processor}}} (\bibinfo {year} {2022}{\natexlab{b}})\BibitemShut
  {NoStop}%
\bibitem [{\citenamefont {Heyl}(2018)}]{heyl_dynamical_2018}%
  \BibitemOpen
  \bibfield  {author} {\bibinfo {author} {\bibfnamefont {M.}~\bibnamefont
  {Heyl}},\ }\bibfield  {title} {\bibinfo {title} {Dynamical quantum phase
  transitions: a review},\ }\href {https://doi.org/10.1088/1361-6633/aaaf9a}
  {\bibfield  {journal} {\bibinfo  {journal} {Reports on Progress in Physics}\
  }\textbf {\bibinfo {volume} {81}},\ \bibinfo {pages} {054001} (\bibinfo
  {year} {2018})}\BibitemShut {NoStop}%
\bibitem [{\citenamefont {Cerezo}\ \emph {et~al.}(2021)\citenamefont {Cerezo},
  \citenamefont {Arrasmith}, \citenamefont {Babbush}, \citenamefont {Benjamin},
  \citenamefont {Endo}, \citenamefont {Fujii}, \citenamefont {McClean},
  \citenamefont {Mitarai}, \citenamefont {Yuan}, \citenamefont {Cincio},\ and\
  \citenamefont {Coles}}]{cerezo2021variational}%
  \BibitemOpen
  \bibfield  {author} {\bibinfo {author} {\bibfnamefont {M.}~\bibnamefont
  {Cerezo}}, \bibinfo {author} {\bibfnamefont {A.}~\bibnamefont {Arrasmith}},
  \bibinfo {author} {\bibfnamefont {R.}~\bibnamefont {Babbush}}, \bibinfo
  {author} {\bibfnamefont {S.~C.}\ \bibnamefont {Benjamin}}, \bibinfo {author}
  {\bibfnamefont {S.}~\bibnamefont {Endo}}, \bibinfo {author} {\bibfnamefont
  {K.}~\bibnamefont {Fujii}}, \bibinfo {author} {\bibfnamefont {J.~R.}\
  \bibnamefont {McClean}}, \bibinfo {author} {\bibfnamefont {K.}~\bibnamefont
  {Mitarai}}, \bibinfo {author} {\bibfnamefont {X.}~\bibnamefont {Yuan}},
  \bibinfo {author} {\bibfnamefont {L.}~\bibnamefont {Cincio}},\ and\ \bibinfo
  {author} {\bibfnamefont {P.~J.}\ \bibnamefont {Coles}},\ }\bibfield  {title}
  {\bibinfo {title} {Variational quantum algorithms},\ }\href
  {https://doi.org/10.1038/s42254-021-00348-9} {\bibfield  {journal} {\bibinfo
  {journal} {Nature Reviews Physics}\ }\textbf {\bibinfo {volume} {3}},\
  \bibinfo {pages} {625} (\bibinfo {year} {2021})}\BibitemShut {NoStop}%
\bibitem [{\citenamefont {Potter}\ and\ \citenamefont
  {Vasseur}(2022)}]{Potter_2022}%
  \BibitemOpen
  \bibfield  {author} {\bibinfo {author} {\bibfnamefont {A.~C.}\ \bibnamefont
  {Potter}}\ and\ \bibinfo {author} {\bibfnamefont {R.}~\bibnamefont
  {Vasseur}},\ }\bibfield  {title} {\bibinfo {title} {Entanglement dynamics in
  hybrid quantum circuits},\ }in\ \href
  {https://doi.org/10.1007/978-3-031-03998-0_9} {\emph {\bibinfo {booktitle}
  {Quantum Science and Technology}}}\ (\bibinfo  {publisher} {Springer
  International Publishing},\ \bibinfo {year} {2022})\ pp.\ \bibinfo {pages}
  {211--249}\BibitemShut {NoStop}%
\bibitem [{\citenamefont {Fisher}\ \emph {et~al.}(2022)\citenamefont {Fisher},
  \citenamefont {Khemani}, \citenamefont {Nahum},\ and\ \citenamefont
  {Vijay}}]{Fisher22}%
  \BibitemOpen
  \bibfield  {author} {\bibinfo {author} {\bibfnamefont {M.~P.~A.}\
  \bibnamefont {Fisher}}, \bibinfo {author} {\bibfnamefont {V.}~\bibnamefont
  {Khemani}}, \bibinfo {author} {\bibfnamefont {A.}~\bibnamefont {Nahum}},\
  and\ \bibinfo {author} {\bibfnamefont {S.}~\bibnamefont {Vijay}},\ }\href
  {https://doi.org/10.48550/ARXIV.2207.14280} {\bibinfo {title} {Random quantum
  circuits}} (\bibinfo {year} {2022})\BibitemShut {NoStop}%
\bibitem [{\citenamefont {Huang}\ \emph {et~al.}(2020)\citenamefont {Huang},
  \citenamefont {Kueng},\ and\ \citenamefont
  {Preskill}}]{huang_predicting_2020}%
  \BibitemOpen
  \bibfield  {author} {\bibinfo {author} {\bibfnamefont {H.-Y.}\ \bibnamefont
  {Huang}}, \bibinfo {author} {\bibfnamefont {R.}~\bibnamefont {Kueng}},\ and\
  \bibinfo {author} {\bibfnamefont {J.}~\bibnamefont {Preskill}},\ }\bibfield
  {title} {\bibinfo {title} {Predicting many properties of a quantum system
  from very few measurements},\ }\href
  {https://doi.org/10.1038/s41567-020-0932-7} {\bibfield  {journal} {\bibinfo
  {journal} {Nature Physics}\ }\textbf {\bibinfo {volume} {16}},\ \bibinfo
  {pages} {1050} (\bibinfo {year} {2020})}\BibitemShut {NoStop}%
\bibitem [{\citenamefont {Webb}(2016)}]{webb_clifford_2016}%
  \BibitemOpen
  \bibfield  {author} {\bibinfo {author} {\bibfnamefont {Z.}~\bibnamefont
  {Webb}},\ }\href {https://doi.org/10.48550/arXiv.1510.02769} {\bibinfo
  {title} {The {Clifford} group forms a unitary 3-design}} (\bibinfo {year}
  {2016})\BibitemShut {NoStop}%
\bibitem [{noa(2022)}]{noauthor_quantinuum_2022}%
  \BibitemOpen
  \href {https://github.com/CQCL/quantinuum-hardware-specifications} {\bibinfo
  {title} {Quantinuum {Hardware} {Specifications}}} (\bibinfo {year}
  {2022})\BibitemShut {NoStop}%
\bibitem [{\citenamefont {Cross}\ \emph {et~al.}(2017)\citenamefont {Cross},
  \citenamefont {Bishop}, \citenamefont {Smolin},\ and\ \citenamefont
  {Gambetta}}]{cross_open_2017}%
  \BibitemOpen
  \bibfield  {author} {\bibinfo {author} {\bibfnamefont {A.~W.}\ \bibnamefont
  {Cross}}, \bibinfo {author} {\bibfnamefont {L.~S.}\ \bibnamefont {Bishop}},
  \bibinfo {author} {\bibfnamefont {J.~A.}\ \bibnamefont {Smolin}},\ and\
  \bibinfo {author} {\bibfnamefont {J.~M.}\ \bibnamefont {Gambetta}},\ }\href
  {https://doi.org/10.48550/arXiv.1707.03429} {\bibinfo {title} {Open {Quantum}
  {Assembly} {Language}}} (\bibinfo {year} {2017})\BibitemShut {NoStop}%
\bibitem [{\citenamefont {Barreiro}\ \emph {et~al.}(2011)\citenamefont
  {Barreiro}, \citenamefont {M{\"u}ller}, \citenamefont {Schindler},
  \citenamefont {Nigg}, \citenamefont {Monz}, \citenamefont {Chwalla},
  \citenamefont {Hennrich}, \citenamefont {Roos}, \citenamefont {Zoller},\ and\
  \citenamefont {Blatt}}]{barreiro2011open}%
  \BibitemOpen
  \bibfield  {author} {\bibinfo {author} {\bibfnamefont {J.~T.}\ \bibnamefont
  {Barreiro}}, \bibinfo {author} {\bibfnamefont {M.}~\bibnamefont
  {M{\"u}ller}}, \bibinfo {author} {\bibfnamefont {P.}~\bibnamefont
  {Schindler}}, \bibinfo {author} {\bibfnamefont {D.}~\bibnamefont {Nigg}},
  \bibinfo {author} {\bibfnamefont {T.}~\bibnamefont {Monz}}, \bibinfo {author}
  {\bibfnamefont {M.}~\bibnamefont {Chwalla}}, \bibinfo {author} {\bibfnamefont
  {M.}~\bibnamefont {Hennrich}}, \bibinfo {author} {\bibfnamefont {C.~F.}\
  \bibnamefont {Roos}}, \bibinfo {author} {\bibfnamefont {P.}~\bibnamefont
  {Zoller}},\ and\ \bibinfo {author} {\bibfnamefont {R.}~\bibnamefont
  {Blatt}},\ }\bibfield  {title} {\bibinfo {title} {An open-system quantum
  simulator with trapped ions},\ }\href {https://doi.org/10.1038/nature09801}
  {\bibfield  {journal} {\bibinfo  {journal} {Nature}\ }\textbf {\bibinfo
  {volume} {470}},\ \bibinfo {pages} {486} (\bibinfo {year}
  {2011})}\BibitemShut {NoStop}%
\bibitem [{h11(2022{\natexlab{b}})}]{h11-data-sheet}%
  \BibitemOpen
  \href@noop {} {\bibinfo {title} {{Quantinuum System Model H1 Product Data
  Sheet}}},\ \bibinfo {howpublished}
  {\url{https://www.quantinuum.com/hardware/h1/},} (\bibinfo {year}
  {2022}{\natexlab{b}})\BibitemShut {NoStop}%
\bibitem [{\citenamefont {Sivarajah}\ \emph {et~al.}(2021)\citenamefont
  {Sivarajah}, \citenamefont {Dilkes}, \citenamefont {Cowtan}, \citenamefont
  {Simmons}, \citenamefont {Edgington},\ and\ \citenamefont
  {Duncan}}]{sivarajah_tket_2021}%
  \BibitemOpen
  \bibfield  {author} {\bibinfo {author} {\bibfnamefont {S.}~\bibnamefont
  {Sivarajah}}, \bibinfo {author} {\bibfnamefont {S.}~\bibnamefont {Dilkes}},
  \bibinfo {author} {\bibfnamefont {A.}~\bibnamefont {Cowtan}}, \bibinfo
  {author} {\bibfnamefont {W.}~\bibnamefont {Simmons}}, \bibinfo {author}
  {\bibfnamefont {A.}~\bibnamefont {Edgington}},\ and\ \bibinfo {author}
  {\bibfnamefont {R.}~\bibnamefont {Duncan}},\ }\bibfield  {title} {\bibinfo
  {title} {t{\textbar}ket : {A} {Retargetable} {Compiler} for {NISQ}
  {Devices}},\ }\href {https://doi.org/10.1088/2058-9565/ab8e92} {\bibfield
  {journal} {\bibinfo  {journal} {Quantum Science and Technology}\ }\textbf
  {\bibinfo {volume} {6}},\ \bibinfo {pages} {014003} (\bibinfo {year}
  {2021})}\BibitemShut {NoStop}%
\bibitem [{\citenamefont {Iqbal}\ \emph {et~al.}(2023)\citenamefont {Iqbal},
  \citenamefont {Tantivasadakarn}, \citenamefont {Gatterman}, \citenamefont
  {Gerber}, \citenamefont {Gilmore}, \citenamefont {Gresh}, \citenamefont
  {Hankin}, \citenamefont {Hewitt}, \citenamefont {Horst}, \citenamefont
  {Matheny}, \citenamefont {Mengle}, \citenamefont {Neyenhuis}, \citenamefont
  {Vishwanath}, \citenamefont {Foss-Feig}, \citenamefont {Verresen},\ and\
  \citenamefont {Dreyer}}]{iqbal_supporting_2023}%
  \BibitemOpen
  \bibfield  {author} {\bibinfo {author} {\bibfnamefont {M.}~\bibnamefont
  {Iqbal}}, \bibinfo {author} {\bibfnamefont {N.}~\bibnamefont
  {Tantivasadakarn}}, \bibinfo {author} {\bibfnamefont {T.}~\bibnamefont
  {Gatterman}}, \bibinfo {author} {\bibfnamefont {J.}~\bibnamefont {Gerber}},
  \bibinfo {author} {\bibfnamefont {K.}~\bibnamefont {Gilmore}}, \bibinfo
  {author} {\bibfnamefont {D.}~\bibnamefont {Gresh}}, \bibinfo {author}
  {\bibfnamefont {A.}~\bibnamefont {Hankin}}, \bibinfo {author} {\bibfnamefont
  {N.}~\bibnamefont {Hewitt}}, \bibinfo {author} {\bibfnamefont
  {C.}~\bibnamefont {Horst}}, \bibinfo {author} {\bibfnamefont
  {M.}~\bibnamefont {Matheny}}, \bibinfo {author} {\bibfnamefont
  {T.}~\bibnamefont {Mengle}}, \bibinfo {author} {\bibfnamefont
  {B.}~\bibnamefont {Neyenhuis}}, \bibinfo {author} {\bibfnamefont
  {A.}~\bibnamefont {Vishwanath}}, \bibinfo {author} {\bibfnamefont
  {M.}~\bibnamefont {Foss-Feig}}, \bibinfo {author} {\bibfnamefont
  {R.}~\bibnamefont {Verresen}},\ and\ \bibinfo {author} {\bibfnamefont
  {H.}~\bibnamefont {Dreyer}},\ }\href {https://zenodo.org/record/7693062}
  {\bibinfo {title} {Supporting {Data} and {Code} for "{Topological} {Order}
  from {Measurements} and {Feed}-{Forward} on a {Trapped} {Ion} {Quantum}
  {Computer}"}} (\bibinfo {year} {2023})\BibitemShut {NoStop}%
\end{thebibliography}%

\small
\newpage

\section{Methods \label{methods}}

\subsection{Entropy measurements from randomised measurements}
\label{methods_entropy}
	
While the deterministic measurement of non-linear quantities, like the entanglement entropy, generally requires full state tomography, recently, robust probabilistic algorithms have been devised ~\cite{van2011power,elben_renyi_2018,vermersch_unitary_2018,brydges_probing_2019,huang_predicting_2020}. These can be used to measure the second-order Renyi entropy
\begin{align}
    S^{(2)} (\rho_A) = - \ln \text{Tr} \rho_A^2.
\end{align}
We follow the protocol from~\cite{brydges_probing_2019} where it is shown that the \emph{purity} of a reduced quantum state can be estimated by
\begin{align}\label{eq:purity}
	\text{Tr} \rho_A^2 = 2^{N_A}\sum_{s_A,s_A'}{(-2)^{D(s_A,s_A')}\overline{P(s_A)P(s_A')}},
\end{align}
where $N_A$ is the subsystem size and $D(s_A,s_A')$ denotes the Hamming distance between the bitstrings corresponding to the computational basis states $s_A$ and $s_A'$. $P(s_A)$ is the probability of measuring $s_A$ in the state $\rho_A$ after applying a random unitary, i.e., $P(s_A)=\langle s_A|U\rho_AU^\dagger|s_A \rangle$, and the overline denotes the average over random unitaries. Importantly, each of the random unitaries is a tensor product of \emph{local} unitaries $u_1\otimes u_2\dots$ which are drawn from the circular unitary ensemble. In principle, one can also choose to simply measure each of the qubits in the $X\text{-},Y\text{-}$~or~$Z\text{-}$basis with equal probability, at the cost of larger statistical fluctuations~\cite{satzinger_realizing_2021}. That is because random Pauli measurements are equivalent to random Clifford gates followed by  computational basis measurements and the Clifford group forms a unitary 3-design~\cite{webb_clifford_2016, huang_predicting_2020}.
	
In the experiment, we compute the purity by averaging over $N_U=72$ random local unitary settings, and for each setting we execute $N_M=256$ shots to estimate the probability distribution $P(s)$. The parameters $N_U$ and $N_M$ have been chosen by using the grid search procedure as described in \cite{brydges_probing_2019} on the emulator of H1-1 ion-trap. We use the unbiased estimator $P(P N_M-1)/(N_M-1)$ for evaluating $P^2(s)$ in \eqref{eq:purity} as described in \cite{vermersch_unitary_2018}. As discussed in the main text, we discard shots where an odd number of anyons was heralded during the error correction. Even without discarding such shots, we get estimates for $\gamma /\ln 2$ of $0.87 \pm 0.055$, and $1.00 \pm 0.090$ for $2\times2$ and $2\times3$ regions respectively. 
	
To evaluate R\'enyi entropies, $S^{(2)}_X$, for each subsystem $X$, we consider different regions of size $2\times2$ and $2\times3$, as shown in the Extended Data Fig.~\ref{ext_fig_ee}(a,b). In Fig.~\ref{ext_fig_ee}(c,d), we show the estimated values of the topological entanglement entropy for each region. The values of $S^{(2)}_X$, as reported in the main text in Fig. \ref{fig_toric_code}(d), have been obtained by taking the mean of subsystem R\'enyi entropies for each region. The error bars for $S^{(2)}_X$ and $\gamma$ have been computed by bootstrapping new samples from the given dataset (which is specified by $U_N=72$ different randomized measurement settings) and then by evaluating the standard deviation of the resulting distribution. 

\subsection{Effects of Measurement Error Mitigation}
While we have not employed any State Preparation and Measurement (SPAM) error mitigation in the main text, we show here the impact of SPAM mitigation on the state preparation procedure.

SPAM error mitigation accounts for the state preparation and readout errors, and in its simplest form, it models the effects of SPAM noise processes on the ideal probability distribution from the quantum device, as
\begin{align}\label{eq_spam}
    P_{\text{noise}} = A^{\otimes n}P_{\text{ideal}},
\end{align}	
where $A$ is the transition matrix that acts locally on each of the $n$ qubits. It is characterized by the probabilities of misreading state $|0\rangle$ as state $|1\rangle$ and vice versa. According to prior characterization of the measurement error in the H1-1 ion-trap, a $\ket{0}$ state on an ion has a $0.1\%$-chance of being read out as $\ket{1}$ and there is a $0.5\%$-chance of $\ket{1}$ erroneously being read as $\ket{0}$ \cite{noauthor_quantinuum_2022}. We can use that information to construct the transition matrix as, 
\begin{align}	
    A = \begin{pmatrix}
        1-0.001 & 0.005\\
        0.001 & 1-0.005
    \end{pmatrix}.
\end{align}	
Note that in \eqref{eq_spam}, we ignore the effects of correlated SPAM errors which occur on multiple sites. We can recover $P_{\text{ideal}}$ by applying $(A^{-1})^{\otimes n}$ on $P_{\text{noise}}$, and this can done efficiently since the action of transition matrix $A$ is local. 	The resulting performance of SPAM error mitigation with and without discarding heralded errors can be seen in  Fig.~\ref{ext_fig_comparisons}(a-d). Without discarding, we get an improvement in energy density from $-0.89$ to $-0.91$, while SPAM correction boosts energy density from $-0.93$ to $-0.95$ using the heralded procedure employed in the main text.

\subsection{Measurement and Decoding of the $A_p$-operator with and without ancillae}

The key advantage of measurement-based over unitary state preparation is the ability to project the state onto an eigenstate of all $A_p$ operators simultaneously. These ``parity check" measurements can be done in one of two ways. More commonly, the parity of the four data qubits on the plaquettes is transferred onto an ancilla using four maximally entangling gates, e.g., to apply the projector $(\mathbb{I} + X^{\otimes 4})/2$, we apply 4 $\text{CNOT}$ gates where the control qubit is an ancilla prepared in $\ket{+}$. It is, however, also possible to obtain constant depth circuits without introducing ancillae. In that case, an ancilla-free parity check needs to be executed, using six two-qubit gates. Both procedures are shown in Extended Data Fig.~\ref{ext_fig_measurement}(a,b). In either case, the state is projected into the even (odd) eigenspace of the stabilizer $X^{\otimes 4})$ upon measuring +1 (-1). However, in the ancilla-free case, we can remove the classically controlled-$Z$ gate to our advantage, as described below. We call the new construction \emph{modified ancilla-free parity check}. The action remains unchanged when the measurement outcome is +1, but, when the outcome is -1, the action of the modified ancilla-free circuit is given by the projector $(\mathbb{I} - X^{\otimes 4})Z_{\text{target}}/2$, where $Z_{\text{target}}$ accounts for the removed classically controlled-$Z$ gate by acting on the target qubit (i.e., the qubit that is measured). The $Z_\mathrm{target}$ autocorrects the measured plaquette at the cost of moving a potential error to an adjacent plaquette. This autocorrection ensures that errors can only accumulate on half of the $X$-type plaquettes and reduces the cost of the subsequent decoding. We emphasize that the use of modified ancilla-free parity checks is done out of convenience - its use is not essential to obtain high-fidelity results.

Since the H1-1 ion-trap is capable of handling up to 20 qubits, using the ancilla-based strategy on four $X$-type plaquettes allows us to save eight two-qubit gates in the state preparation circuit for free, and those plaquettes are shown in Fig.~\ref{ext_fig_measurement}(c) by hatching with slanted(\textbackslash) lines. The remaining four $X$-type plaquettes are measured by using the modified ancilla-free measurement circuit and they are shown in Fig.~\ref{ext_fig_measurement}(c) by hatching with crossed ($\times$) lines, the red arrows pointing to the target qubits. With this construction, whenever we measure -1 (indicating error or the presence of anyon in the plaquette) on the modified ancilla-free plaquette, the additional $Z_{\text{target}}$ action moves the error/anyon into the diagonally adjacent plaquette in which the arrow is pointing.
		
To remove the anyon pairs, we have implemented a simple lookup-table decoder which handles each of the $2^4/2 = 8$ error possibilities explicitly. We have implemented a decoder that is compatible with OpenQASM 2.0 for the all the results that we present in the main text~\cite{cross_open_2017}. OpenQASM 2.0 does not support conditioning on individual bits in the classical register and this causes a substantial increase in the number of classically conditioned single-qubit gates during the correction. Later in the development, we also considered a more optimal decoder (i.e., with significantly less conditional gates in total and asymptotically linear cost in system size). The resulting energy densities achieved by the former and later decoders are given in Fig.~\ref{ext_fig_comparisons}(b) and Fig.~\ref{ext_fig_comparisons}(e) respectively. We find that both decoders give almost the same energy density. We also test the case where every $X$-type plaquette is prepared by measurement with an ancilla circuit (Fig.~\ref{ext_fig_measurement}(a)), measuring four stabilizers using available ancillae and then reuse those ancillae to measure remaining $X$-plaquettes. Although this implementation requires $4\times(6-4)=8$ fewer two-qubit gates, reuse of ancilla qubits in this fashion generally increases the circuit depth, execution time, and potential for memory errors. The results for this procedure are given in Fig.~\ref{ext_fig_comparisons}(f) showing a slight improvement in the energy density, in particular for the $X$-type plaquettes, since the noise in the ancilla-based preparation strategy is biased towards corrupting $Z$-type plaquettes (cf. main text and Extended Data Figure~\ref{ext_fig_measurement}).

We also consider a decoder where we do not measure one of the $X^{\otimes 4}$ stabilizers. Since in the noiseless case, anyonic excitations occur in pairs, we can deduce the state of unmeasured $X$-type plaquette from the parity of measured stabilizers and apply the corresponding error correction steps. Expectation values of stabilizers for this decoder are shown Fig.~\ref{ext_fig_comparisons}(g).

\label{sec_measurement_and_decoding}

\subsection{State Preparation and dynamics of the model with a defect}

The state preparation strategy for the defective model proceeds similarly to the defect-free state. Since the model requires 15 data qubits, we have five leftover qubits on the 20 qubit H1-1 ion-trap which we use as ancilla qubits to prepare the $X$-type plaquettes 1, 3 and 4 and the two defect-plaquettes. The remaining three $X$-type plaquettes (cf. Fig.~\ref{fig_defect}(a)) are prepared using the modified ancilla-free parity check circuits (cf. section~\ref{sec_measurement_and_decoding}). Again, we use a simple lookup-table decoder to remove the errors while exploiting the fact that errors never occur in plaquettes which are prepared by modified ancilla-free measurement.

While in the defect-free case only two settings are necessary to measure all stabilizers using single-qubit measurements, for the defective case, we use four settings to obtain one observation for all stabilizers. We measure plaquettes $(0,2,5,7,14)$, $(3,5,10,11,13)$, $(1,3,4,6,11)$, and $(0,6,8,12,14)$ in the first, second, third, and fourth setting respectively. This splitting has the advantage that each measurement contributes to the expectation value and the corner and defect plaquettes are measured twice (cf. Fig.~\ref{fig_defect}(a)).
	
For the anyon transmutation, the sequence $X_{12} X_{13} Z_{6} Z_{5}$ has been applied after the state preparation (cf. Fig.~\ref{fig_defect}(c)). After each of the single-qubit gates, the corner and the defect stabilizers are measured 1200 times and all other plaquettes are measured 600 times. For the anyon interferometry, the circuit $H_{\text{anc}} Y_{10} \text{CZ}_{7} \text{CZ}_{4} \text{CZ}_{8} \text{CZ}_{10} Y_{10}\ket{+}_\text{anc} \otimes \ket{\text{gs}}$ has been applied where the control qubit is the ancilla. The real part of the braiding phase is the measured $\braket{Z}$ expectation value of the ancilla at the end of this sequence.

We have also examined the transmutation of a magnetic anyon into an electric anyon while utilizing quantum nondemolition (QND) measurements of stabilizers \cite{barreiro2011open}. This has the advantage that one can measure the whole transmutation trajectory in one shot, while not destroying the anyon. We construct the circuit such that it begins by preparing the toric code with defects. Then we apply $X_{12}$, which creates a pair of flux anyons and measure the stabilizers on plaquettes 1, 4, 6, 8 and 12 using the circuit shown in Fig.~\ref{ext_fig_measurement}(a). The resulting measurement outcomes for these stabilizers are given in Fig.~\ref{ext_fig_nd_trans}(left). Then, we apply $X_{13}$, to move one of the flux anyon into the defect plaquette and repeat the QND measurement of the same stabilizers (see Fig.~\ref{ext_fig_nd_trans}(middle) for the resulting expectation values). Finally, we apply $Z_6$ which moves the anyon out of the defect plaquette and again do a QND measurement of the same stabilizers as above. Afterward, we measure all other stabilizers destructively. The results of these measurements are given in Fig.~\ref{ext_fig_nd_trans}(right).
 
\subsection{Circuit Construction, Gate Count and Error Budget}

The native gate set of H1-1 ion-trap consists of the single-qubit gates
\begin{align*} 
U_{1q}\left(\theta = \left\{\frac{\pi}{2},\pi \right\}, \phi \right) &= e^{-i(\cos \phi X +\sin \phi Y )\theta/2}, \\ 
R_z(\lambda) &=e^{-iZ \lambda/2}
\end{align*}
and the arbitrary-angle entangling gate $RZZ(\theta) = e^{-i \theta/2 Z\otimes Z}$ \cite{h11-data-sheet}. Specifications at the time of the experiment (November and December 2022) indicated average two-qubit gate fidelity of $99.7\%$, one-qubit fidelity of $99.996\%$, state preparation and measurement fidelity of $99.6\%$ and memory-error per depth-1 circuit time per qubit of $1-99.97\%$.

The state preparation circuit for toric code including the decoder requires $484$ one qubit gates and $4\times4+6\times 4 = 40$ two-qubit gates. In the case of toric code with defects, after compilation into native gate set, the circuit contains $423$ one-qubit gates and $3\times4+5\times 2+6\times3 = 40$ two-qubit gates. The circuits were compiled to the native gate set and sent to the device using TKET~\cite{sivarajah_tket_2021}.

We estimate the global fidelity with the target state using  the same randomized measurement data set that we also use to compute topological entanglement entropies (cf. section~\ref{methods_entropy}) by using the framework of \emph{shadow density matrices} \cite{huang_predicting_2020}. We construct shadow density matrix for each randomized measurement setting, take its overlap with the target wavefunction and then calculate the mean value. We report a global fidelity with the $\braket{Z^\text{hori}} = \braket{Z^\text{vert}} = 1$ toric code ground state of $\braket{\text{gs} |\rho_\text{prepared}|  \text{gs}} = 0.80 \pm 0.049$.

Multiplying the gate error of all 40 two-qubit gates ($0.997^{40} \approx 0.887$), 484 one-qubit gates ($0.99996^{484} \approx 0.981$) the memory error on all of the 20 qubits accumulating during 6 depth-1 circuit times ($0.9997^{6 \times 20} \approx 0.965$) and the state preparation and measurement error on 20 + 4 (reused) qubits ($0.996^{24} \approx 0.9082$) leads to a global damping factor of approximately $0.762$ which is compatible with the estimated global fidelity.

\section*{Data availability}
The numerical data that support the findings of this study, including a full list of shots is available on the Zenodo repository~\cite{iqbal_supporting_2023}.

\section*{Code availability} The code used for quantum circuit construction, submission and data analysis is available on the Zenodo repository~\cite{iqbal_supporting_2023}.

\section*{Acknowledgements}
This work was made possible by a large group of people, and the authors would like to thank the entire Quantinuum team for their many contributions. We are greatful for helpful discussions and feedback from Ciaran Ryan-Anderson, Konstantinos Meichanetzidis, Ben Criger, Eli Chertkov, Kevin Hemery, Ramil Nigmatullin, Reza Haghshenas, Khaldoon Ghanem, Alexander Schuckert, Ella Crane, David Hayes, and Natalie Brown. N.T. is supported by the Walter Burke Institute for Theoretical Physics at Caltech. R.V. is supported by the Harvard Quantum Initiative Postdoctoral Fellowship in Science and Engineering. A.V. is supported by NSF-DMR 2220703 and A.V. and R.V. are supported by the Simons Collaboration on Ultra-Quantum Matter, which is a grant from the Simons Foundation (618615, A.V.). The experimental data in this work was produced by the Quantinuum H1-1 trapped ion quantum computer, Powered by Honeywell. H.D. acknowledges support by the German Federal Ministry of Education and Research (BMBF) through the project EQUAHUMO (grant number 13N16069) within the funding program quantum technologies - from basic research to market.

\section*{Author contributions}
M.I. wrote the code generating the circuits and submitted all experiments. The data analysis was done by M.I. and H.D.
N.T., R.V., and A.V. contributed to the ideation, theory and experiment design. M.F. contributed to the theory, including the decoder and the characterisation of device noise.
T.M.G., J.A.G, K.G., D.G., A.H., N.H., C.V.H., M.M., T.M. and B.N. operated the ion-trap during the experiment.
H.D. drafted the manuscript, to which all authors contributed.


\section*{Additional information}
Correspondence and requests for materials should
be addressed to H.D.

\clearpage
\onecolumngrid

\section{Extended Data Figures}

\subsection{Covariances and Noise Bias}
\begin{figure}[!h]
    \centering
    \includegraphics[width=\textwidth,scale=1]{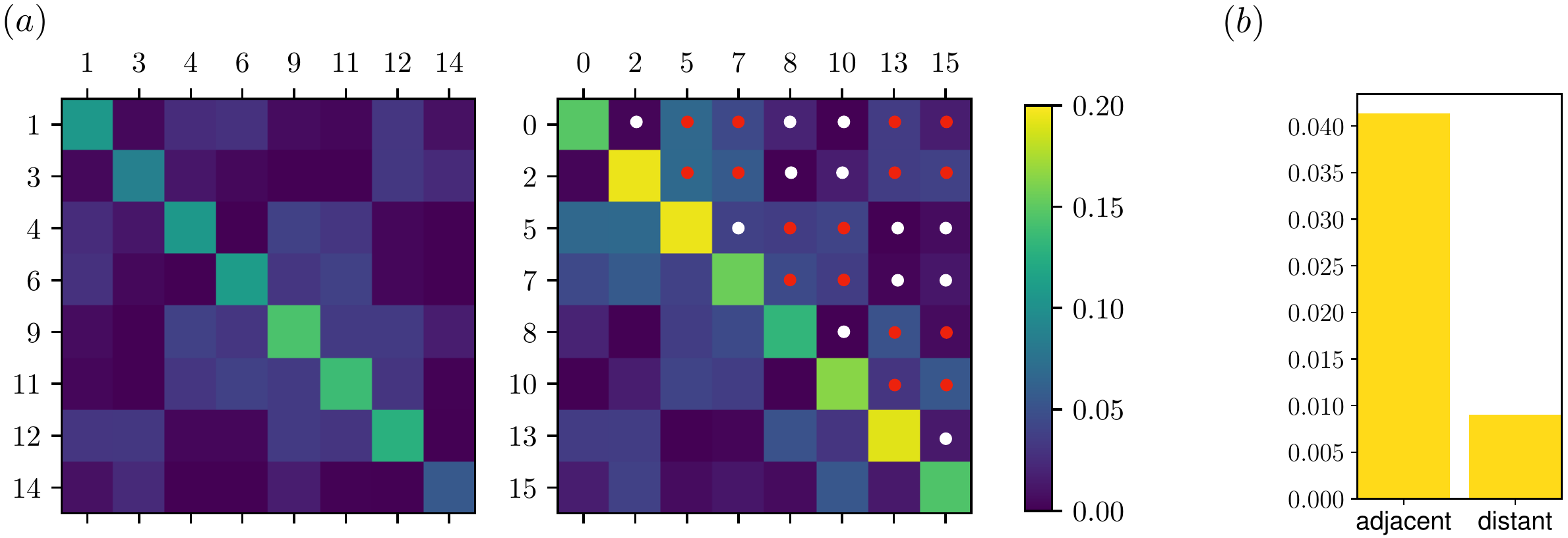}
    \caption{\textbf{Additional data confirming the gate noise bias towards $Z$-errors}. A single $Z$-flip occuring during the $(\mathbb{I}+X^{\otimes 4})/2$ projection flips two adjacent $Z$-plaquettes but leaves $X$-plaquettes invariant. (a) Covariances of plaquettes $\braket{P_p P_q} - \braket{P_p}\braket{P_q}$, $X$-plaquettes on the left ($P=A$), $Z$-plaquettes on the right ($P=B$). Neighbouring (non-neighbouring) plaquettes are marked by a red (white) dot. (b) Average $Z$-plaquette correlation functions $\braket{B_p B_q}- \braket{B_p}\braket{B_q}$ over all nearest-neighbour ('adjacent') and non-nearest-neighbour ('distant') plaquettes.}
    \label{ext_fig_noise_bias}
\end{figure}
		
\subsection{Topological Entanglement Entropy}
\begin{figure}[!h]
    \centering
    \includegraphics[width=52em,scale=1]{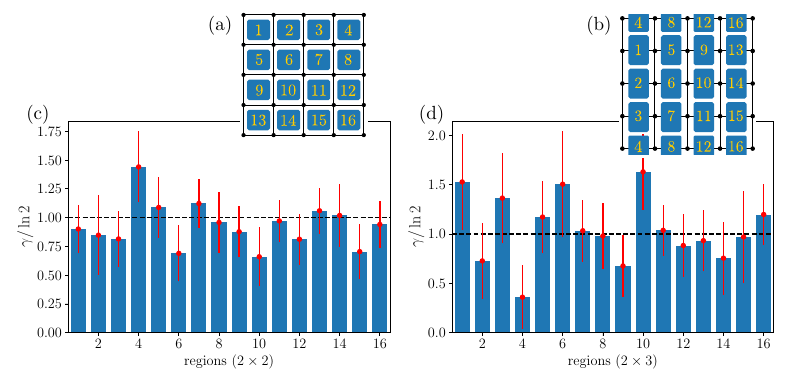}
    \caption{\textbf{Additional data for the entanglement entropy measurements.} (a, b) $2\times2$ and $2\times3$ regions (solid rectangles) and their labels on $4 \times 4$ torus. Each $2\times 3$ region consists of two vertically adjacent plaquettes. (c, d) Entanglement entropy correction $\gamma$ for each region using the protocol discussed in the main text and sec.~\ref{methods_entropy}. The maximum value of the error bars is $\pm 0.35\ (\pm 0.54)$ for $2\times2$ ($2\times3$) regions. In the $2\times2$ case, the value of $\gamma$ is averaged over all rotations.} \label{ext_fig_ee}
\end{figure}

\subsection{Parity check circuits and stabilizer measurement layout}
\begin{figure}[!h]
    \centering
    \includegraphics[width=50em,scale=1]{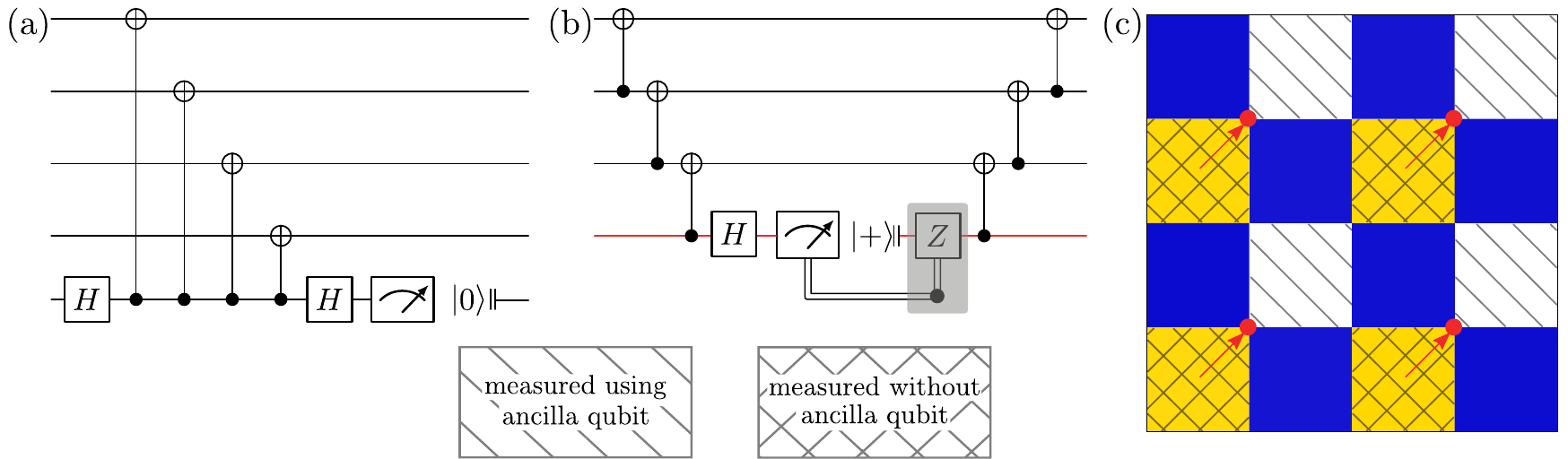}
    \caption{\textbf{Parity check protocols.} (a) Circuits to measure $X^{\otimes 4}$ stabilizers with an ancilla. CNOT gates are compiled to $H_\mathrm{target} \mathrm{CZ} H_\mathrm{target}$ in the device. Native two-qubit gates are slightly more likely to cause a $Z$-error than an $X$-error. Such a $Z$ error propagates through the enclosing Hadamard gate to become an $X$-error which causes two excitations on the $Z$-type plaquettes which are adjacent to the corrupted qubit. (b) Circuits to measure $X^{\otimes 4}$ stabilizers without ancilla qubit. We use modified ancilla-free parity check circuits in our implementation in which the conditional-$Z$ (shaded) is removed (cf. section~\ref{sec_measurement_and_decoding}). (c) $4\times4$ torus showing measured plaquettes. Plaquettes hatched by slanted lines (\textbackslash) are measured using ancilla qubit circuits, and the plaquettes where we employ modified ancilla-free measurement circuit are hatched by crossed lines ($\times$). Arrowheads point at the target qubit of the ancilla-free protocol. The twelve colored plaquettes are stabilized after the projection step, while the remaining four empty plaquettes require feed-forward correction. 
     \label{ext_fig_measurement}}
\end{figure}

\subsection{Non-demolition anyon tracing}
\begin{figure}[!h]
    \centering
    \includegraphics[width=1.0\textwidth]{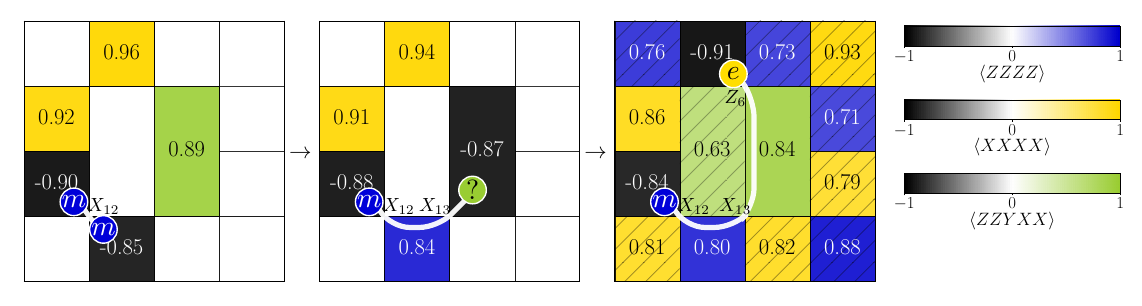}
    \caption{\textbf{Transmutation of a magnetic into an electric anyon using non-demolition measurements.} In the main text, all qubits are measured destructively after each step of the transmutation. Alternatively, at the cost of introducing extra gates, a full transmutation can be observed in a single shot. Plaquettes hatched by slanted (/) lines are measured destructively.}
    \label{ext_fig_nd_trans}
\end{figure}
\clearpage

\subsection{Energy densities with SPAM error mitigation and for strictly ancilla-based measurement circuits}
\begin{figure}[!h]
    \centering
    \includegraphics[width=1.0\textwidth]{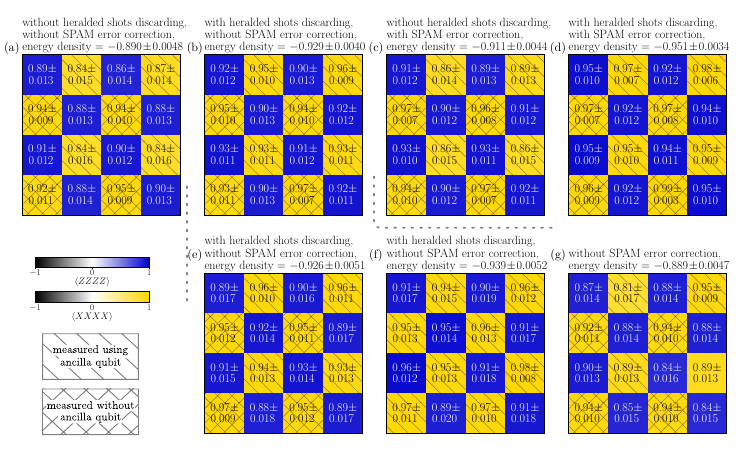}
    \caption{\textbf{Comparison of the variation of the preparation strategy described under Methods.} (a-d) Effect of discarding heralded shots and state preparation and measurement (SPAM) error mitigation. (e) Energy density for the optimized decoder. (f) Using ancilla qubits for all stabilizer measurements. (g) Strategy where one plaquette is not projected but instead inferred from the overall constraint that there must be an even number of anyons.}
\label{ext_fig_comparisons}
\end{figure}
\clearpage

\end{document}